\documentclass[final,leqno,onefignum,onetabnum]{siamltex1213}

\usepackage{amsmath}
\usepackage{amssymb}
\usepackage{enumerate}

\newtheorem{remark}{Remark}[section]

\begin{document}

\title{CONVERGENCE AND ERROR PROPAGATION RESULTS ON A LINEAR ITERATIVE UNFOLDING METHOD}

\author{ANDR\'AS L\'ASZL\'O\thanks{Wigner Research Centre for Physics, Konkoly-Thege M.u. 29-33, Budapest, H-1121, Hungary.
(e-mail: \email{laszlo.andras@wigner.mta.hu}).}}

\maketitle
\slugger{sinum}{xxxx}{xx}{x}{x--x}

\begin{abstract}
Unfolding problems often arise in the context of statistical data 
analysis. Such problematics occur when the probability 
distribution of a physical quantity is to be measured, but it is randomized 
(smeared) by some well understood process, such as a non-ideal detector 
response or a well described physical phenomenon. In such case 
it is said that the original probability distribution of interest is folded 
by a known response function. The reconstruction of the original probability 
distribution from the measured one is called 
unfolding. That technically involves evaluation of the non-bounded inverse 
of an integral operator over the space of $L^1$ functions, which is known to 
be an ill-posed problem. For the pertinent regularized operator inversion, 
we propose a linear iterative formula and provide proof of convergence in 
a probability theory context. 
Furthermore, we provide formulae for error estimates at finite iteration 
stopping order which are of utmost importance in practical applications: 
the approximation error, the propagated statistical error, 
and the propagated systematic error can be quantified. 
The arguments are based on the Riesz-Thorin theorem mapping the original $L^1$ 
problem to $L^2$ space, and subsequent application of ordinary $L^2$ spectral 
theory of operators. A library implementation in C of the algorithm along with 
corresponding error propagation is also provided. A numerical example also 
illustrates the method in operation.
\end{abstract}



\pagestyle{myheadings}
\thispagestyle{plain}
\markboth{Andr\'as L\'aszl\'o}{Convergence and error propagation results on a linear iterative unfolding method}

\section{Introduction}

In analysis of experimental data one commonly faces the problem that the probability density 
function (\textit{pdf}) of a given physical quantity of interest is to be measured, 
but some random physical process, such as the intrinsic behavior of the 
measurement apparatus smears it. The reconstruction of the pertinent unknown 
pdf of interest based on the observed smeared pdf and on the known response 
function of the measurement procedure is called unfolding.

More specifically, one of the most common unfolding scenarios turning up 
in experimental data analysis is the following. 
Let $x\mapsto f(x)$ be the unknown pdf which we intend to reconstruct, 
$(y,x)\mapsto \rho(y\vert x)$ be the known response function of the smearing effect, 
and we assume that $y\mapsto g(y) = \int \rho(y\vert x)\,f(x)\,\mathrm{d}x$ 
is the measured pdf after the smearing effect, called folding. In practice, actually often only 
a statistical estimator of $g$ can be measured. Or, putting it differently, 
$g$ often contains an additional error term $y\mapsto e(y)$ originating from statistical 
counting and unaccounted systematic measurement distortions, in which case 
one has $y\mapsto g(y) = \int \rho(y\vert x)\,f(x)\,\mathrm{d}x \,+\, e(y)$ 
as the measured pdf estimator. The task of unfolding is to 
provide some close estimate for $x\mapsto f(x)$, given $y\mapsto g(y)$ and 
$(y,x)\mapsto \rho(y\vert x)$ along with 
some estimate on $y\mapsto e(y)$, i.e.\ to solve the above linear integral 
equation. It is quite well known in the literature that such a problem is 
numerically ill-posed. The primary reason for this is Banach's closed graph 
theorem: due to the pertinent theorem a generic folding operator maps certain 
distant pdfs to close ones whose differences after the folding are shadowed 
by the contribution of the measurement error term $e$. That quite well 
understood phenomenon is summarized e.g.\ in
\cite{laszlo2012, laszlo2006, cowan2002, blobel2008, zech2010, kuusela2015, kuusela2016, dembinski2013}.

The problematics of unfolding can also be formulated using a language possibly 
more familiar to statisticians \cite{dattner2011, dattner2016}. Let 
$x_{1},\dots,x_{n}$ be statistical instances of a probability variable $x$, 
i.e.\ independent identically distributed random variables, 
each having the same but unknown pdf $f$. In the experimental setting, 
merely the random variables $y_{i}=x_{i}+\varepsilon_{x_{i},i}$ ($i=1,\dots,n$) 
are observed, i.e.\ the original $x_{i}$ ($i=1,\dots,n$) random variables 
corrupted by an $x$-dependent, but otherwise independent identically distributed 
error variable $\varepsilon_{x}$, having a known $x$-dependent pdf 
$\varepsilon_{x}\mapsto\rho(\varepsilon_{x} +x|x)$ for each 
fixed value of $x$ as a condition. Given all these, the task of unfolding is
to provide an estimator for the pdf $f$ of the undistorted probability variable $x$. 
In some real experimental situation, it also happens that the individual 
observed samples $y_{i}=x_{i}+\varepsilon_{x_{i},i}$ ($i=1,\dots,n$) are not 
published, only their pdf estimator $g$ is made available, for instance because there is some 
correction procedure on the pdf level, e.g.\ for inefficiencies. Also, our 
model $(y,x)\mapsto \rho(y\vert x)$ for the response function 
might be systematically inaccurate, for which inaccuracy only an upper bound might 
be known. Therefore, often not the sample based observational model, but 
rather the previously discussed pdf estimator based observational model is 
more practical to handle. But whichever way the problem is formulated 
--- based on individual samples or on pdfs --- the task remains to be ill-posed.

In order to overcome the ill-posedness of the unfolding problem, all the methods 
use restrictions on the unknown pdf, and in some special cases properties of 
the response function can also be used to improve the situation. For instance, 
in the field of image or signal processing, the shape of the response function 
is translationally invariant in an exact manner, i.e.\ for all $x,y,z$ one has 
$\rho(y\vert x+z)=\rho(y-z\vert x)$, and thus the unfolding reduces to the 
problematics of deconvolution. In the language of statistical samples, this 
would correspond to the observational model when $y_{i}=x_{i}+\varepsilon_{i}$ 
($i=1,\dots,n$) are observed, with independent identically distributed random 
variables $\varepsilon_{i}$ of a known distribution, not depending on $x$. 
Due to the applicational importance of the special case of deconvolution problems, that branch has a whole stream of 
literature \cite{dattner2011, dattner2016, fan1991, hesse2006, lacour2013, liu1989, stefanski1990, kalifa2003}. 
The statistical deconvolution methods heavily rely on the applicability of 
convolution theorem for the Fourier transformed pdfs, which is possible due to the translational 
invariance of the shape of the response function, i.e.\ relies on the fact that 
the probability variables $\varepsilon_{i}$ ($i=1,\dots,n$) are independent 
identically distributed and are independent from $x$. The ill-posedness of the problem, 
similarly to the case of any generic unfolding method, is regularized by finding and approximative solution. The optimal 
approximation is controlled by the application of the minimax principle: 
for a given estimate of the true deconvolved pdf, a loss (penalty) function 
is defined, and the minimum of the worst case expected loss is looked for as 
a function of the regularization parameters. It is worth to note that most of 
the advanced statistical deconvolution methods can work on unbinned samples, 
i.e.\ they do not need an a priori histograming of the observed data. In 
Section \ref{numericalexample} an illustrative numerical unfolding toy model 
application is presented, which also tries to clarify that in an experimental 
context more general approaches than deconvolution are also needed in order 
to handle real measurement situations.

Also in the case of generic --- i.e.\ non-deconvolution --- unfolding problems a regularization method must be 
applied \cite{laszlo2012, cowan2002, blobel2008, zech2010, kuusela2015, kuusela2016, dembinski2013, kalifa2003, hoecker1996, dagostini1995, zech2013} 
and an approximate solution of the folding integral equation within 
a reduced set of allowed pdfs is searched for. The approximation is controlled by 
some regularization parameters whose particular value brings in a certain degree 
of arbitrariness to the unfolded pdf (approximation error), which is often difficult to quantify. 
There are basically three main widespread ways in the literature addressing the problem of 
regularization.
\begin{enumerate}[(i)]
\item In certain data analysis problems a parametric ansatz for the unknown pdf 
$f$ is justified. In that case, one can construct the folded version of $f$ 
by the response function $\rho$ numerically, and that can be fitted to the 
observed folded pdf $g$, for instance via a maximum likelihood method. Such 
method is used for instance in inclusive particle identification in experimental 
high energy particle physics (see for instance \cite{laszlo2008}). Due to the 
ill-posedness of the unfolding problem, one may run into a situation in particular 
cases when the fit is insensitive to some details of the parametrically given 
$f$. In other words: the log-likelihood function ($\chi^{2}$) may be flat 
in the direction of certain parameters of the ansatz for $f$.
\item Bin-by-bin fitting of the histogramed $f$, such that when numerically 
folding it by $\rho$ the result gets close to the observed folded pdf $g$, e.g.\ in 
a maximum likelihood sense. This is very similar to approach (i) with 
every bin amplitude of the histogramed $f$ being a fit parameter. This method 
is basically equivalent to the naive inversion of the discretized folding 
operator as a matrix. Due to the ill-posedness of the unfolding problem, this 
is not satisfactory in itself. The usual procedure is to add some artificial 
penalty function to the log-likelihood function ($\chi^{2}$) in order to suppress 
the large local gradients. If that is performed, the method can deliver meaningful 
answers, but the introduced systematic bias by the additional penalty function 
is difficult to quantify. In addition, similarly to the method (i), the fit can 
be slightly insensitive to the details of $f$ due to the ill-posedness of the 
problem. The so called SVD methods \cite{hoecker1996} are implementations of this idea.
\item There are also iterative methods which intend to approximate the true 
pdf $f$, given the measured folded pdf $g$ and the response function $\rho$. 
One of the most popular and most promising methods is the method of convergent 
weights, also called iterative Bayesian unfolding. It was first discovered and 
applied by Richardson \cite{richardson1972} and Lucy \cite{lucy1974} for image 
processing. Later it was re-discovered and applied to tomography problems by 
Shepp and Vardi \cite{shepp1982}, and by Kondor \cite{kondor1983}. The first 
serious mathematical scrutiny of the method was done by M\"ulthei and Schorr 
\cite{multhei1987mma, multhei1987nim}. In the mid-90s d'Agostini re-discovered 
and popularized the algorithm in the high energy physics community 
\cite{dagostini1995}. Recently, Zech \cite{zech2013} studied possible optimal 
iteration stopping criteria for the algorithm. One of the main advantages of 
the method of convergent weights or Bayesian unfolding is, that it takes into 
account the non-negativity and the unitness of the integral of the true pdf 
$f$ in an exact manner. Furthermore, if the measured folded pdf $g$ was a 
histogram, i.e.\ its values fluctuate according to Poisson counting statistics, 
then the iterative approximants to $f$ have increasing likelihood 
\cite{multhei1987mma}, i.e.\ the algorithm is a realization of a 
maximum-likelihood approximation. Most unfortunately, despite of the research 
efforts \cite{multhei1987mma}, there are no results stating that the method 
is convergent, although numerical evidence suggests its convergent nature. 
Moreover, there are no exact error propagation formulae available.
\end{enumerate}
\vspace*{2mm}

In case of a consistent method the approximation error should converge to 
zero when the regularization parameters are relaxed. 
In case of an iterative method, an approximating sequence 
$\left(f_{N}\right)_{N\in\mathbb{N}_{0}}$ to the unknown $f$ is constructed 
and the regularization parameter is merely the iteration stopping order 
$N_{\mathrm{max}}$, i.e.\ a threshold index in the approximating sequence. 
When an iterative unfolding method is consistent, 
the approximation error, i.e.\ the distance of $f_{N}$ to the true unknown $f$ 
must converge to zero with increasing number of iterations $N$. 
Although the above consistency property 
is an obvious minimal requirement for any unfolding method, often this is not 
easy to show analytically.

In a previous paper \cite{laszlo2012} we proposed a linear iterative unfolding 
method, discussed its pros and cons in comparison to other techniques, 
provided detailed description from the practical point of view for 
experimentalists, along with providing a set of relevant application examples. 
In the present paper we provide formal mathematical proofs for the claims 
therein for the proposed unfolding method:
\begin{enumerate}[(i)]
\item proof of consistency, i.e.\ that the approximation error 
converges to zero with increasing number of iterations, 
\item explicit formula for the approximation error at finite iteration order,
\item explicit formula for the propagated statistical errors on the unfolded 
pdf at finite iteration order given the statistical errors of the measured folded pdf,
\item explicit formula for the propagated systematic errors on the unfolded 
pdf at finite iteration order given the systematic errors of the 
measured folded pdf or of the response function.
\end{enumerate}
Because of (ii)--(iv) the competing error terms become calculable, and 
therefore these can be used to define an optimal iteration stopping criterion. 
In addition, the pertinent error terms can be determined at this optimum. 
The quantification of these are of utmost importance when presenting unfolded experimental results, 
and is generally an unresolved task for other widely used unfolding methods. 
The key mathematical ingredient of the proofs are mapping our originally 
$L^{1}$ problem to the $L^{2}$ space using Riesz-Thorin theorem, and 
using spectral representation of the operators therein. The actual iteration 
formula is formally motivated by a preconditioned Neumann-Landweber-Richardson series, 
but these are not automatically convergent in case of $L^{1}$ problems: 
our specific preconditioning makes the iteration convergent in the $L^{1}$ 
setting, given some quite generic conditions. The proposed method also does not 
rely on an inherent discretization of the pdfs: it does work also in the 
continuum limit or with any type of density estimators.\footnote{Some unfolding methods rely on an 
inherent discretization of pdfs in the problem, and use the assumed 
discretization as an implicit regularization. Our method does not use such trick.}

The obtained results can be particularly interesting as the proposed method 
can be considered as the ``linearized'' version of the method of convergent 
weights or iterative Bayesian unfolding \cite{dagostini1995, zech2013, richardson1972, lucy1974, shepp1982, kondor1983, multhei1987mma, multhei1987nim}. 
By understanding the convergence conditions and error propagation for the 
proposed method, the studies of M\"ulthei and Schorr \cite{multhei1987mma} 
could eventually be completed on the Bayesian iteration, which would be a 
significant improvement in the field.

The paper is organized as follows: in Section \ref{unfoldinproblem} 
the problem of unfolding is introduced in a mathematically rigorous way, 
and the basic properties of generic folding operators are discussed. 
In Section \ref{lineariterativeunfolding} our 
proposed unfolding method is introduced and proofs are provided for its 
above listed properties. In Section \ref{measures} we generalize a bit our results for the case of 
probability measures which are not described by pdfs.
In Section 
\ref{discretecase} we restrict our results to the special case when the 
unfolding problem is discrete: this presentation may be better understood 
by statisticians or experimental physicists not specialized in functional analysis.
In Section \ref{numericalexample} a concrete numerical example is shown.
Finally, in Section \ref{conclusions} we summarize.

\section{Mathematical properties of folding operators and the unfolding}
\label{unfoldinproblem}

In the text we shall abbreviate by \textit{pdf} the notion of probability 
density function, by \textit{cpdf} the notion of conditional probability 
density function. We shall rely on the usual terminology in functional 
analysis and measure theory \cite{lax2002, rudin1973}. 
As such, the notion of Lebesgue almost everywhere or Lebesgue 
almost every, shall be abbreviated by \textit{a.e.}

Let $X$ and $Y$ be finite dimensional real vector spaces equipped with the 
Lebesgue measure --- unique up to a global positive 
normalization factor. Let $L^{1}(X)$ and $L^{1}(Y)$ denote the Banach 
spaces of $X\rightarrow \mathbb{C}$ and $Y\rightarrow \mathbb{C}$ Lebesgue 
integrable function equivalence classes, respectively, where the equivalence of functions 
is defined by being a.e.\ equal. As usual in functional analysis 
texts, we shall call these function equivalence classes simply functions. 
We shall also use the notion of essential bound for such a function which 
is the smallest upper bound valid a.e.

\vspace*{2mm}
\begin{definition}
Let $\rho:\,Y\times X\rightarrow \mathbb{R}_{0}^{+},\,(y,x)\mapsto\rho(y\vert x)$ be a cpdf over the 
product space $Y\times X$, i.e.\ a non-negative Lebesgue measurable function 
which satisfies $\forall x\in X:\,\int\rho(y\vert x)\,\mathrm{d}y = 1$. Then, 
the linear operator
\begin{eqnarray}
A_{\rho}\;:\;L^{1}(X)\rightarrow L^{1}(Y),\;(x\mapsto f(x))\,\mapsto\,\left(y\mapsto\int\rho(y\vert x)\,f(x)\,\mathrm{d}x\right)
\label{folding}
\end{eqnarray}
is called the folding operator by $\rho$, where the function $\rho$ is called 
the response function of the folding.
\end{definition}

\vspace*{2mm}
\begin{remark}
The following basic properties of folding operators are direct consequences 
of the definition.
\begin{enumerate}[(i)]
\item A possible usual generalization of the notion of folding operator is when 
inefficiencies are also allowed, i.e.\ the less restrictive condition
$\forall x\in X:\,\int\rho(y\vert x)\,\mathrm{d}y \leq 1$ is required for 
the response function $\rho$ of the folding operator $A_{\rho}$. 
The results throughout the paper are also valid for that case.
\item By Fubini's theorem, a folding is a well defined linear operator.
\item It is also quite evident \cite{laszlo2006} that such operator is 
continuous in the $L^{1}$ operator norm (i.e.\ in probabilistic sense), 
moreover $\Vert A_{\rho}\Vert_{L^{1}(X)\rightarrow L^{1}(Y)}=1$, while $\Vert A_{\rho}\Vert_{L^{1}(X)\rightarrow L^{1}(Y)}\leq1$ whenever 
inefficiencies are allowed.
\end{enumerate}
\end{remark}

It is seen that such a folding operator $A_{\rho}$ is quite well 
behaved: it is linear and is continuous in the probabilistic sense, i.e.\ close 
pdfs are mapped to close pdfs in the $L^{1}$ sense \cite{laszlo2012}.

A quite important class of folding operators are convolutions, in which case 
the shape of the response function is translationally invariant.

\vspace*{2mm}
\begin{definition}
A folding operator $A_{\rho}$ is called convolution whenever the response 
function $\rho$ is translationally invariant in the sense that
$Y=X$ and $\forall x,y,z\in X\,:\,\rho(y\vert x+z)=\rho(y-z\vert x)$.
\end{definition}

\vspace*{2mm}
\begin{remark}
The following properties of convolution operators are well-known results \cite{laszlo2006, arfken2013, bracewell1999}.
\begin{enumerate}[(i)]
\item In case a folding operator $A_{\rho}$ is a convolution, 
the response function $\rho$ may be expressed by the single pdf 
$\eta:=\rho(\cdot\vert 0)$ in the form 
$\forall x,y\in X:\,\rho(y\vert x)=\eta(y-x)$. The alternative 
notation $\eta\star f:=A_{\rho}f$ is often used in such case ($f\in L^{1}(X)$). 
Note that convolution is commutative, i.e.\ one has 
$\eta\star f = f\star \eta$ for all $\eta,f\in L^{1}(X)$.
\item A convolution operator is not onto, and its image is not closed.
\item The image of a convolution operator is dense if and only if the Fourier 
transform of the convolver function is nowhere zero (Wiener's approximation theorem).
\item A convolution operator is one-to-one if and only if the Fourier transform 
of the convolver function is a.e.\ nonzero.
\item Consequently, the inverse of a convolution operator, whenever exists, cannot 
be continuous. This is because a convolution is everywhere defined on the closed 
set $L^{1}(X)$, it is continuous, and therefore it has closed graph by Banach's 
closed graph theorem; but since the inverse operator's domain is not closed, 
again by Banach's closed graph theorem, it cannot be continuous.
\end{enumerate}
\end{remark}

\vspace*{2mm}
Since the convolution operators form a quite large example class of folding 
operators, we can state that a generic folding operator's inverse, whenever 
exists, is not continuous. This finding is often referred to as: the inversion of a 
generic folding operator is ill-posed. The argument goes as follows: we have an unknown 
pdf $f$, a known response function $\rho$, and a measured pdf $g=A_{\rho}f+e$ 
where $e$ represents a small measurement error term. Then, when one would set 
$A_{\rho}^{-1}g=f+A_{\rho}^{-1}e$, the error term $e$ contains modes not 
in the domain of $A_{\rho}^{-1}$ in which case $A_{\rho}^{-1}e$ is not 
meaningful, or when approximated numerically, this term shall diverge. 
Note that even if all modes of $e$ were in the domain of $A_{\rho}^{-1}$, 
the smallness of $A_{\rho}^{-1}e$ is not guaranteed even though $e$ is small. 
The ill-posedness of a generic unfolding problem may also be stated as: 
if $f_{1}$ and $f_{2}$ are distant pdfs, then $g_{1}:=A_{\rho}f_{1}+e_{1}$ and 
$g_{2}:=A_{\rho}f_{2}+e_{2}$ may be close pdfs, i.e. we lose discrimination 
power on pdfs after a folding \cite{laszlo2012}. The presented argument also 
warns us against relying solely on the so called \textit{closure test} when verifying 
an unfolding algorithm: whenever some unfolding method gives some estimate 
$\hat{f}$ for the unknown pdf $f$, it is usually argued that 
$A_{\rho}\hat{f}\approx A_{\rho}f$ confirms the validity of the estimate 
$\hat{f}$. Clearly, in the light of our observations this is not enough, 
as $\hat{f}$ may be still far from $f$ in the probabilistic distance.

Due to the ill-posedness of the unfolding problem, any unfolding method needs 
to use some kind of regularization: some assumption on the original (unknown) 
pdf, and a way to search for an approximative solution depending on some regularization 
parameters. Furthermore, the convergence 
to the original pdf when relaxing these parameters can usually 
be only achieved in some weak sense, not in the probabilistic norm of 
$L^{1}(X)$. The most commonly applied unfolding strategies are summarized in 
\cite{laszlo2012, cowan2002, blobel2008, zech2010, kuusela2015, kuusela2016, dembinski2013, kalifa2003, hoecker1996, dagostini1995, zech2013}.

\section{A linear iterative unfolding method}
\label{lineariterativeunfolding}

Since the folding equation Eq.(\ref{folding}) is linear, it is quite natural 
to try applying some iterative inversion methods known in functional 
analysis, when approximating the true solution $f$. 
One such self-suggesting method is Neumann series 
\cite{lax2002, rudin1973} which guarantees that 
whenever for a continuous linear operator $A$ over a Banach space one has 
$\Vert I-A\Vert<1$ ($I$ being the identity operator), then 
$A^{-1}=\sum_{n=0}^{\infty}(I-A)^{n}$ where the convergence holds in the 
operator norm. That convergence requirement, however, 
cannot be satisfied in case of a probability theory folding operator 
because for such an 
operator one has $\Vert I-A_{\rho}\Vert_{L^{1}\rightarrow L^{1}} = 2$ as shown in \cite{laszlo2006}. 
The Richardson iteration, based on similar requirements, does not work for the same reason. 
An other evident choice would be the Landweber iteration \cite{landweber1951} 
known in the theory of Fredholm integral equations \cite{lax2002, rudin1973}. 
This assumes, in first place, that 
the unknown function $f$ and the result of the folding $g$ resides in the space 
of square integrable functions $L^{2}(X)$, furthermore that the response function 
$\rho$ satisfies the regularity condition $\int\int\big\vert\rho(y\vert x)\big\vert^{2}\,\mathrm{d}y\,\mathrm{d}x < \infty$. 
The latter regularity condition, unfortunately, is violated in case of a 
generic cpdf, on the contrary to the common belief in the literature.\footnote{It is evidently seen that this 
regularity condition does not hold for any convolution. It is also seen at the 
price of some calculation that this situation cannot be repaired by a 
compactification mapping, i.e.\ if we map the support set of our pdfs and 
response function into a compact region of $Y$ and $X$.}

Despite of the fact that neither the Neumann series, nor the Richardson iteration, 
nor the Landweber iteration can be directly applied to an unfolding problem, they provide a possible 
starting point. Motivated by these algorithms we proposed a linear iterative 
unfolding method for a probability theory context, i.e.\ for the $L^{1}$ space \cite{laszlo2012}. 
The section is continued by recalling notions necessary for studying the pertinent algorithm.

In the followings we shall denote by $L^{p}(X)$ the Banach space of 
$X\rightarrow\mathbb{C}$ functions \cite{lax2002, rudin1973} which are 
Lebesgue integrable of the $p$-th power ($1\leq p \leq \infty$). 
The special case $L^{\infty}(X)$ for $p=\infty$ 
is defined as the Banach space of the $X\rightarrow\mathbb{C}$ essentially 
bounded functions with their norm being the essential bound.

\vspace*{2mm}
\begin{remark} The argumentation in the followings relies on some known results.
\begin{enumerate}[(i)]
\item
The Riesz-Thorin theorem 
\cite{folland1999} states 
that if $1\leq q\leq r\leq\infty$ and $F\subset L^{q}(X)\cap L^{r}(X)$ is a dense 
linear subspace in both $L^{q}(X)$ and $L^{r}(X)$, furthermore a linear operator 
$T:F\rightarrow L^{q}(X)\cap L^{r}(X)$ is bounded both in the $L^{q}(X)$ and 
$L^{r}(X)$ norm, then for all $q\leq p\leq r$ values $F\subset L^{p}(X)$, it 
is dense in $L^{p}(X)$, $T[F]\subset L^{p}(X)$ and $T$ is bounded in the 
$L^{p}(X)$ norm. Thus, $T$ is uniquely extendable as an 
$L^{p}(X)\rightarrow L^{p}(X)$ bounded linear operator. In addition we have that
\begin{eqnarray}
\left\Vert T\right\Vert_{L^{p}\rightarrow L^{p}} \leq \max\left( \left\Vert T\right\Vert_{L^{q}\rightarrow L^{q}}, \left\Vert T\right\Vert_{L^{r}\rightarrow L^{r}} \right)
\label{rieszthorin}
\end{eqnarray}
holds for the operator norms.
\item
An important consequence of the Riesz-Thorin theorem is that a convolution 
operator $\eta\star(\cdot)$ by a function $\eta\in L^{1}(X)$ is well defined 
and continuous in $L^{p}(X)$ for all $1\leq p\leq\infty$ and its operator 
norm is bounded by $\Vert \eta\Vert_{L^{1}}$. This obviously holds for the 
$p=1$ and $p=\infty$ case due to H\"older's inequality, and then it is implied 
for all $1< p<\infty$ as well by the pertinent theorem. As a consequence, using 
the commutativity of convolution, it also 
follows that if $\varphi\in L^{p}(X)$ and $\eta\in L^{1}(X)$ then 
$\varphi\star \eta\in L^{p}(X)$, i.e.\ pdfs may be mapped into $L^{p}(X)$ via 
convolution by pdfs integrable on the $p$-th power.
\item
We shall use in the followings the spectral representation \cite{rudin1973} 
of normal operators over complex separable Hilbert spaces. Let $T$ be a normal operator 
over the pertinent space, i.e.\ a densely defined linear operator with closed 
graph, satisfying $T^{*}T=TT^{*}$, $(\cdot)^{*}$ being the adjoint. Then there 
exists a unique projection valued measure $P$ over the Borel sets of the 
spectrum set of $T$, $\mathrm{Sp}(T)$, such that
\begin{eqnarray}
T=\int_{\lambda\in\mathrm{Sp}(T)} \lambda\,\mathrm{d}P(\lambda)
\label{spectral1}
\end{eqnarray}
holds, where the integral is defined in the weak sense. That is, for all elements 
$f,g$ in the Hilbert space one has a complex valued Borel measure 
$\left<f,P(\cdot)g\right>$ such that
\begin{eqnarray}
\left<f, T g\right>=\int_{\lambda\in\mathrm{Sp}(T)} \lambda\;\mathrm{d}\left<f,P(\lambda)g\right>.
\label{spectral2}
\end{eqnarray}
In addition, one has that if $M$ is a polynomial, then $M(T)$ is also normal operator, 
furthermore
\begin{eqnarray}
M(T)=\int_{\lambda\in\mathrm{Sp}(T)} M(\lambda)\,\mathrm{d}P(\lambda)
\label{spectral3}
\end{eqnarray}
is satisfied in the same sense.
\end{enumerate}
\label{rieszthorincor}
\end{remark}

\vspace*{2mm}
Throughout the argumentation we will need the notion of transpose folding 
which is introduced below.

\vspace*{2mm}
\begin{definition}
If $A_{\rho}$ is a folding operator such that the response function 
$\rho(\cdot\vert x)$ is 
square-integrable for all $x\in X$, then for all $k\in L^{2}(Y)$ 
the expression
\begin{eqnarray}
A_{\rho}^{T}k :=\left(x\mapsto\int k(y)\,\rho(y\vert x)\,\mathrm{d}y\right)
\label{transposefolding}
\end{eqnarray}
is meaningful and defines a linear map from $L^{2}(Y)$ to the Lebesgue 
measurable functions $X\rightarrow \mathbb{C}$. We call the linear operator 
$A_{\rho}^{T}$ the transpose folding.
\end{definition}

\subsection{\textbf{\textit{The iterative approximation}}}

Equipped with the listed notions, we can introduce the following 
approximating sequence for solution of the unfolding problem. Let $g=A_{\rho}f$ 
be our unfolding problem where $f$ is to be determined, with $g$ and $\rho$ 
being known. We try to approximate the solution in the form:
\begin{eqnarray}
K_{\rho} & := & \sup_{x\in X} \int \int \rho(y\vert z)\, \rho(y\vert x)\, \mathrm{d}y\, \mathrm{d}z, \cr
f_{0}    & := & K_{\rho}^{-1} A_{\rho}^{T} g, \cr
f_{N+1}  & := & f_{N} + \left(f_{0} - K_{\rho}^{-1}A_{\rho}^{T}A_{\rho}f_{N}\right) \cr
         &    & (N\in \mathbb{N}_{0}).
\label{iterativesolution}
\end{eqnarray}
This is, formally, the iterative expression for Neumann series after 
preconditioning by $K_{\rho}^{-1} A_{\rho}^{T}$, i.e.\ for the composite 
operator $K_{\rho}^{-1} A_{\rho}^{T}A_{\rho}$.

\subsection{\textbf{\textit{Convergence conditions}}}

The following theorem shows that under quite generic conditions the approximating 
sequence $\left(f_{N}\right)_{N\in\mathbb{N}_{0}}$ in terms of Eq.(\ref{iterativesolution}) 
is well-defined and converges 
to $f$ whenever $A_{\rho}$ is one-to-one, and it converges to the closest possible 
function to $f$ whenever $A_{\rho}$ is not one-to-one.

\vspace*{2mm}
\begin{theorem} (Convergence)
Let $A_{\rho}$ be a folding operator and assume that its response function 
$\rho$ has the property that for all $x\in X$ the function $\rho(\cdot|x)$ 
is square-integrable, furthermore $K_{\rho}<\infty$. 
Assume that the unknown pdf $f$ in the unfolding 
problem $g=A_{\rho}f$ is square-integrable. Then:
\begin{enumerate}[(i)]
\item For any compact set $U\subset X$:
\begin{eqnarray}
\lim_{N\rightarrow\infty} \frac{1}{\mathrm{Volume}(U)} \int_{x\in U} \left(f-\mathcal{P}_{\mathrm{Ker}(A_{\rho})}f-f_{N}\right)(x) \,\mathrm{d}x = 0,
\label{convergence1}
\end{eqnarray}
where $\mathcal{P}_{\mathrm{Ker}(A_{\rho})}$ is the $L^{2}$ orthogonal projection onto 
the kernel set of $A_{\rho}$.
\item We have that
\begin{eqnarray}
\lim_{N\rightarrow\infty} \left\Vert f-\mathcal{P}_{\mathrm{Ker}(A_{\rho})}f-f_{N} \right\Vert_{L^{2}} = 0
\label{convergence2}
\end{eqnarray}
and the convergence is monotone.
\end{enumerate}
\label{convergence}
\end{theorem}
\begin{proof}
It is seen that whenever the regularity condition $\forall x\in X:\,\rho(\cdot|x)\in L^{2}(Y)$ holds, 
the function
\begin{eqnarray}
\alpha\,:\; X\times X\rightarrow\mathbb{R}^{+}_{0}, \;(z,x)\mapsto\alpha(z,x) := \int \rho(y|z) \rho(y|x) \,\mathrm{d}y
\end{eqnarray}
is well defined. By construction, it is symmetric, i.e.\ 
$\forall z,x\in X:\,\alpha(z,x) = \alpha(x,z)$. Furthermore, because of $K_{\rho}<\infty$ and symmetricity,
\begin{eqnarray}
\sup_{x\in X}\int_{z\in X} \alpha(z,x) \,\mathrm{d}z = \sup_{z\in X}\int_{x\in X} \alpha(z,x) \,\mathrm{d}x = K_{\rho} < \infty
\end{eqnarray}
holds. With this, we see that the operator $A_{\rho}^{T}A_{\rho}$ is well 
defined as $L^{1}(X)\rightarrow L^{1}(X)$ and is bounded, its $L^{1}\rightarrow L^{1}$ operator norm being $K_{\rho}$. 
This is because for any $f\in L^{1}(X)$
\begin{eqnarray}
\left\Vert A_{\rho}^{T}A_{\rho}f \right\Vert_{L^{1}} = \int \left\vert \int \alpha(z,x)f(x)\,\mathrm{d}x \right\vert \,\mathrm{d}z \cr
 \qquad \leq \int \int \alpha(z,x)\left\vert f(x)\right\vert\,\mathrm{d}x \,\mathrm{d}z = \int \left(\int \alpha(z,x) \,\mathrm{d}z\right) \left\vert f(x)\right\vert\,\mathrm{d}x \cr
 \qquad \leq \sup_{x\in X} \left(\int_{z\in X} \alpha(z,x) \,\mathrm{d}z\right) \,\int_{x\in X}\left\vert f(x)\right\vert\,\mathrm{d}x = K_{\rho} \left\Vert f\right\Vert_{L^{1}}
\end{eqnarray}
due to of monotonicity of integration, Fubini's theorem and H\"older's inequality. 
It is also seen that the operator $A_{\rho}^{T}A_{\rho}$ is well defined as 
$L^{\infty}(X)\rightarrow L^{\infty}(X)$ and is bounded, its $L^{\infty}\rightarrow L^{\infty}$ 
operator norm being $K_{\rho}$. That is because for any $f\in L^{\infty}(X)$
\begin{eqnarray}
\left\Vert A_{\rho}^{T}A_{\rho}f \right\Vert_{L^{\infty}} = \sup_{z\in X} \left\vert \int \alpha(z,x)f(x)\,\mathrm{d}x \right\vert \cr
 \qquad \leq \sup_{z\in X} \int \alpha(z,x)\left\vert f(x)\right\vert \,\mathrm{d}x \leq \sup_{z\in X} \left(\int \alpha(z,x) \,\mathrm{d}x \sup_{x\in X} \left\vert f(x)\right\vert\right) \cr
 \qquad = \sup_{z\in X} \left(\int_{x\in X} \alpha(z,x) \,\mathrm{d}x\right) \,\sup_{x\in X}\left\vert f(x)\right\vert = K_{\rho} \left\Vert f\right\Vert_{L^{\infty}}
\end{eqnarray}
due to monotonicity of integration and H\"older's inequality.

Now, using Riesz-Thorin theorem 
we have that the operator $A_{\rho}^{T}A_{\rho}$ is well-defined as 
$L^{2}(X)\rightarrow L^{2}(X)$ and is bounded, 
its $L^{2}\rightarrow L^{2}$ operator norm being bound by $K_{\rho}$. 
It is also easily seen that for any $f\in L^{2}(X)$ one has 
$\left<f,A_{\rho}^{T}A_{\rho}f\right> = \left<A_{\rho}f,A_{\rho}f\right>\geq 0$, 
therefore it is a self adjoint and positive operator in $L^{2}(X)$. 
Thus, its spectrum lies within the interval $[0,K_{\rho}]$. For brevity, 
we introduce the notation $A:=K_{\rho}^{-1}A_{\rho}^{T}A_{\rho}$ for the 
re-normalized composite folding operator.

Let us observe that the iterative formula Eq.(\ref{iterativesolution}) may also 
be written in the series expansion form $f_{N}=\sum_{n=0}^{N}(I-A)^{n}f_{0}$ 
where we have that $f_{0}=A f$, $f$ being the unknown pdf. This form 
is particularly useful because then we see by induction that 
$\sum_{n=0}^{N}(I-A)^{n}A = I-(I-A)^{N+1}$, i.e.\ we have the explicit formula 
$f-f_{N}=(I-A)^{N+1}f$ for the residual term.

By the observed properties of $A$ it is quite evident that $\mathrm{Sp}(A)\subset[0,1]$.
Thus, there exists a unique projection valued measure $P$ on the Borel sets 
of $[0,1]$ such that 
\begin{eqnarray}
A = \int_{\lambda\in[0,1]} \lambda \,\mathrm{d}P(\lambda)
\end{eqnarray}
in the weak sense. This implies that for any $h\in L^{2}(X)$ we have
\begin{eqnarray}
\left<h,f-f_{N}\right> & = & \int_{\lambda\in[0,1]} (1-\lambda)^{N+1} \,\mathrm{d}\left<h,P(\lambda)f\right> \cr
                       & = & \int_{\lambda\in\{0\}} (1-\lambda)^{N+1} \,\mathrm{d}\left<h,P(\lambda)f\right> + \cr
                       &   & \int_{\lambda\in]0,1]} (1-\lambda)^{N+1} \,\mathrm{d}\left<h,P(\lambda)f\right>.
\end{eqnarray}
Since $\int_{\lambda\in\{0\}} (1-\lambda)^{N+1} \,\mathrm{d}P(\lambda) = \mathcal{P}_{\mathrm{Ker}(A_{\rho})}$ for all $N\in\mathbb{N}_{0}$, 
we arrive at the identity
\begin{eqnarray}
\left<h,f-\mathcal{P}_{\mathrm{Ker}(A_{\rho})}f-f_{N}\right> = \int_{\lambda\in]0,1]} (1-\lambda)^{N+1} \,\mathrm{d}\left<h,P(\lambda)f\right>,
\end{eqnarray}
and by the monotonicity of integration
\begin{eqnarray}
\left\vert\left<h,f-\mathcal{P}_{\mathrm{Ker}(A_{\rho})}f-f_{N}\right>\right\vert \leq \int_{\lambda\in]0,1]} \left\vert 1-\lambda\right\vert^{N+1} \,\mathrm{d}\left\vert\left<h,P(\lambda)f\right>\right\vert
\label{upperbound}
\end{eqnarray}
also holds, where the symbol $\vert\cdot\vert$ when applied to complex valued measures 
denotes variation, which is analogous to absolute value of complex valued functions.
The measure $\left<h,P(\cdot)f\right>$ on $[0,1]$ has finite variation 
and the function sequence $\lambda\mapsto(1-\lambda)^{N+1}$ ($N\in\mathbb{N}_{0}$) 
is bounded independently of $N$ and converges pointwise to zero on $]0,1]$, therefore by Lebesgue's 
theorem of dominated convergence \cite{lax2002, rudin1973} we have that the 
sequence of integrals converges to zero. Thus, the first part of the theorem is 
proved by setting $h:=\frac{1}{\mathrm{Volume}(U)}\,\chi_{{}_{U}}$.

The second part of the theorem is proved by observing that
\begin{eqnarray}
\left\Vert f - \mathcal{P}_{\mathrm{Ker}(A_{\rho})}f - f_{N} \right\Vert_{L^{2}}^{2} & = & \left<f,\left((I-A)^{N+1}-\mathcal{P}_{\mathrm{Ker}(A_{\rho})}\right)^{2}f\right>\cr
 & = & \int_{\lambda\in]0,1]} (1-\lambda)^{2N+2} \,\mathrm{d}\left<f,P(\lambda)f\right>
\end{eqnarray}
where $\left<f,P(\cdot)f\right>$ is a non-negative valued finite measure 
and the integrand which is also non-negative, 
has a bound independent of $N$, furthermore it 
monotonically decreases at each point to zero with increasing $N$. 
Therefore, by Lebesgue's theorem of dominated convergence 
and by the monotonicity of integration we have that the pertinent expression 
converges to zero with increasing $N$ in a monotonically decreasing way.
\end{proof}

\vspace*{2mm}
\begin{remark}
The following remarks clarify the meaning of Theorem~\ref{convergence} in the 
context of a probability theory setting.
\begin{enumerate}[(i)]
\item For any folding operator $A_{\rho}$ the response function may be 
conditioned to have the regularity condition 
$\forall x\in X:\,\rho(\cdot|x)\in L^{2}(X)$ by 
convolving it with a square-integrable 
pdf $\eta$ whose Fourier transform is nowhere vanishing. Namely, one can solve 
the modified problem $\eta\star g = A_{\eta\star\rho}f$ for $f$ 
instead of the original form $g = A_{\rho}f$. In that way, the 
transpose folding operator can always be made well-defined.
When such a treatment is applied, the iteration modifies as
\begin{eqnarray}
K_{\eta\star\rho} & := & \sup_{x\in X} \int \int (\eta\star\rho)(y\vert z)\, (\eta\star\rho)(y\vert x)\, \mathrm{d}y\, \mathrm{d}z, \cr
f_{0}    & := & K_{\eta\star\rho}^{-1} A_{\eta\star\rho}^{T} \;\eta\star g, \cr
f_{N+1}  & := & f_{N} + \left(f_{0} - K_{\eta\star\rho}^{-1}A_{\eta\star\rho}^{T}A_{\eta\star\rho}f_{N}\right) \cr
         &    & (N\in \mathbb{N}_{0}).
\label{iterativesolution2}
\end{eqnarray}
with the very same convergence properties as in the previous theorem.
\item The regularity condition $K_{\rho}<\infty$ (or $K_{\eta\star\rho}<\infty$) holds 
for a quite large class of response functions in a probability theory context. 
Namely, it is easy to check that if $A_{\rho}$ is a convolution, then 
$K_{\rho}=1$. For other practical cases, this condition may be checked 
numerically as done in \cite{laszlo2012}. It is shown e.g.\ that for the 
response function of particle energy measurement with a typical calorimeter 
device, one has $K_{\rho}\approx 1.4$. Also the response function of particle 
momentum measurement using bending in magnetic field has the pertinent 
regularity property.
\item The regularity condition for the unknown pdf $f$, i.e.\ that it has 
to be square-integrable, holds for a quite generic class of pdfs. This 
is automatic for instance for any pdf which is known to be essentially bounded.
\item When the convergence condition is satisfied, it is seen that if 
$A_{\rho}$ is one-to-one, the approximating functions 
$\left(f_{N}\right)_{N\in\mathbb{N}_{0}}$ converge to the original unknown pdf 
$f$. When $A_{\rho}$ is not one-to-one, then $\left(f_{N}\right)_{N\in\mathbb{N}_{0}}$ 
converge to the closest possible function $f-\mathcal{P}_{\mathrm{Ker}(A_{\rho})}f$.
\item The meaning of convergence result (i) in the context of probability theory 
is that the approximating functions $\left(f_{N}\right)_{N\in\mathbb{N}_{0}}$ converge in the 
sense that the probability of each compact set $U\subset X$ is restored to the 
maximum possible extent, but 
the rate of convergence might be different for different sets. When the pdfs 
are measured or modeled by histograms, as usual in statistical data processing, 
this means binwise convergence of the restored histograms, the convergence rate 
being possibly different for different histogram bins. The more global 
convergence result (ii) does not have a direct probability 
theory interpretation, but shall have a role in the estimation of approximation error 
at finite iteration order $N$.
\item Note that whenever our pdfs are modeled by histograms, 
the operation of histogram binning may also be regarded as 
part of the folding operator as described in \cite{laszlo2012}, and thus it 
is wise to include its effect in the folding operator $A_{\rho}$. This might be 
done for instance by modeling the true (unknown) pdf $f$ and its iterative 
approximates $f_{N}$ as histograms binned on much wider domain with larger 
binning density than the measured pdf $g$. In such approximation 
the folding operator $A_{\rho}$ may be thought of as a real matrix which is 
not square.
\end{enumerate}
\label{improvement}
\end{remark}

\subsection{\textbf{\textit{Estimation of approximation error}}}

The convergence result means that the residual term (approximation error) 
$f-\mathcal{P}_{\mathrm{Ker}(A_{\rho})}f-f_{N}$ of the approximating 
sequence defined by Eq.(\ref{iterativesolution}) decreases to zero 
with increased iteration order $N$ in the sense that it decreases to 
zero when averaged over any compact set, i.e.\ we have binwise 
convergence in the language of histograms. However, it would be very useful to 
quantify the approximation error at finite $N$ in order to define some 
stopping criterion. To achieve this, we need to recall a result 
from the theory of projection valued measures.

\vspace*{2mm}
\begin{remark}
Let P be a projection valued measure of some separable Hilbert space 
over the Borel sets of $\mathbb{C}$. Then, whenever 
$\alpha$ and $\beta$ are $\mathbb{C}\rightarrow\mathbb{C}$ measurable functions, 
while $h$ and $f$ are elements of the Hilbert space, one has
\begin{eqnarray}
\left\vert \int_{\lambda\in\mathbb{C}} \alpha(\lambda)\,\beta(\lambda)\,\mathrm{d}\left<h,P(\lambda)f\right>\right\vert \cr
\qquad \leq \sqrt{\int_{\lambda\in\mathbb{C}} |\alpha(\lambda)|^{2} \,\mathrm{d}\left<h,P(\lambda)h\right>} \sqrt{\int_{\lambda\in\mathbb{C}} |\beta(\lambda)|^{2} \,\mathrm{d}\left<f,P(\lambda)f\right>}
\end{eqnarray}
and the same inequality also holds when $\alpha$ and $\beta$ are interchanged \cite{rudin1973}. 
This upper bound is in the analogy of the Cauchy-Schwarz inequality.
\label{generalizedcauchyschwarz}
\end{remark}

\vspace*{2mm}
The following theorem helps to quantify the approximation error at a finite iteration order $N\in\mathbb{N}_{0}$.

\vspace*{2mm}
\begin{theorem} (Approximation error) Take the iterative solution for the unfolding problem as in Eq.(\ref{iterativesolution}) and 
assume that the convergence conditions of Theorem~\ref{convergence} hold. 
Then, the distance of an $N$-th iterate $f_{N}$ from the closest possible 
function to the true unfolded pdf $f$ in the average over a compact set 
$U\subset X$ has the following upper bounds:
\begin{enumerate}[(i)]
\item One has
\begin{eqnarray}
\left\vert\frac{1}{\mathrm{Volume}(U)}\int_{x\in U} \left( f-\mathcal{P}_{\mathrm{Ker}(A_{\rho})}f - f_{N} \right)(x)\,\mathrm{d}x\right\vert \cr
\qquad \leq \frac{1}{\sqrt{\mathrm{Volume}(U)}}\,\left\Vert f-\mathcal{P}_{\mathrm{Ker}(A_{\rho})}f - f_{N} \right\Vert_{L^{2}}.
\end{eqnarray}
\item Similarly, when $\mathrm{Ker}(A_{\rho})$ is not projected out:
\begin{eqnarray}
\left\vert\frac{1}{\mathrm{Volume}(U)}\int_{x\in U} \left( f - f_{N} \right)(x)\,\mathrm{d}x\right\vert \cr
\qquad \leq \frac{1}{\sqrt{\mathrm{Volume}(U)}}\,\left\Vert f - f_{N} \right\Vert_{L^{2}}.
\end{eqnarray}
\item
In addition,
\begin{eqnarray}
\left\vert\frac{1}{\mathrm{Volume}(U)}\int_{x\in U} \left( f-\mathcal{P}_{\mathrm{Ker}(A_{\rho})}f - f_{N} \right)(x)\,\mathrm{d}x\right\vert \cr
\qquad \leq \left\Vert f-\mathcal{P}_{\mathrm{Ker}(A_{\rho})}f\right\Vert_{L^{2}} \, \left\Vert \xi_{{}_{U}}-\mathcal{P}_{\mathrm{Ker}(A_{\rho})}\xi_{{}_{U}} - \xi_{{}_{U,N}}\right\Vert_{L^{2}}
\end{eqnarray}
is valid, where $\xi_{{}_{U}}:=\frac{1}{\mathrm{Volume}(U)}\chi_{{}_{U}}$ and $\xi_{{}_{U,N}}$ is 
the $N$-th iterative approximation of $\xi_{{}_{U}}$ in terms of Eq.(\ref{iterativesolution}). Namely, 
$\xi_{{}_{U,0}}:=K_{\rho}^{-1} A_{\rho}^{T}\xi_{{}_{U}}$ and $\xi_{{}_{U,N+1}}:=\xi_{{}_{U,N}} + \left(\xi_{{}_{U,0}} - K_{\rho}^{-1}A_{\rho}^{T}A_{\rho}\xi_{{}_{U,N}}\right)$.
\item Similarly, one has
\begin{eqnarray}
\left\vert\frac{1}{\mathrm{Volume}(U)}\int_{x\in U} \left( f - f_{N} \right)(x)\,\mathrm{d}x\right\vert \cr
\qquad \leq \left\Vert f \right\Vert_{L^{2}} \, \left\Vert \xi_{{}_{U}} - \xi_{{}_{U,N}}\right\Vert_{L^{2}}
\end{eqnarray}
when $\mathrm{Ker}(A_{\rho})$ is not projected out.
\item The identity
\begin{eqnarray}
\left\vert\frac{1}{\mathrm{Volume}(U)}\int_{x\in U} \left( f - f_{N} \right)(x)\,\mathrm{d}x\right\vert \cr
\qquad = \left\vert \int \left(\xi_{{}_{U}}-\xi_{{}_{U,N}}\right)(x)\,f(x) \,\mathrm{d}x\right\vert
\end{eqnarray}
also holds.
\end{enumerate}
\label{approximationerror}
\end{theorem}
\begin{proof}
These are direct consequence of spectral representation of the operator 
$A:=K_{\rho}^{-1}A_{\rho}^{T}A_{\rho}$ as in the proof of Theorem~\ref{convergence} 
from which
\begin{eqnarray}
\left|\left<h,f-\mathcal{P}_{\mathrm{Ker}(A_{\rho})}f - f_{N}\right>\right|  =  \left|\int_{\lambda\in]0,1]} 1\, (1-\lambda)^{N+1} \,\mathrm{d}\left<h,P(\lambda)f\right>\right| \cr
\;\; \leq \sqrt{\int_{\lambda\in ]0,1]} \left|1\right|^{2} \,\mathrm{d}\left<h,P(\lambda)h\right>} \, \sqrt{\int_{\lambda\in ]0,1]} \left|(1-\lambda)^{N+1}\right|^{2} \,\mathrm{d}\left<f,P(\lambda)f\right>} \cr
\end{eqnarray}
and
\begin{eqnarray}
\left|\left<h,f-\mathcal{P}_{\mathrm{Ker}(A_{\rho})}f - f_{N}\right>\right|  =  \left|\int_{\lambda\in]0,1]} 1\, (1-\lambda)^{N+1} \,\mathrm{d}\left<h,P(\lambda)f\right>\right| \cr
\;\; \leq \sqrt{\int_{\lambda\in ]0,1]} \left|1\right|^{2} \,\mathrm{d}\left<f,P(\lambda)f\right>} \, \sqrt{\int_{\lambda\in ]0,1]} \left|(1-\lambda)^{N+1}\right|^{2} \,\mathrm{d}\left<h,P(\lambda)h\right>} \cr
\end{eqnarray}
follows with arbitrary $h\in L^{2}(X)$. These may be rewritten as:
\begin{eqnarray}
\left|\left<h,f-\mathcal{P}_{\mathrm{Ker}(A_{\rho})}f - f_{N}\right>\right| \cr
 \qquad \leq \left\Vert h-\mathcal{P}_{\mathrm{Ker}(A_{\rho})}h \right\Vert_{L^{2}} \left\Vert \left((I-A)^{N+1}-\mathcal{P}_{\mathrm{Ker}(A_{\rho})}\right)f \right\Vert_{L^{2}}
\end{eqnarray}
and
\begin{eqnarray}
\left|\left<h,f-\mathcal{P}_{\mathrm{Ker}(A_{\rho})}f - f_{N}\right>\right| \cr
 \qquad \leq \left\Vert f-\mathcal{P}_{\mathrm{Ker}(A_{\rho})}f \right\Vert_{L^{2}} \left\Vert \left((I-A)^{N+1}-\mathcal{P}_{\mathrm{Ker}(A_{\rho})}\right)h \right\Vert_{L^{2}}.
\end{eqnarray}
Then by using the fact that 
$\left((I-A)^{N+1}-\mathcal{P}_{\mathrm{Ker}(A_{\rho})}\right)f = f-\mathcal{P}_{\mathrm{Ker}(A_{\rho})}f - f_{N}$ and 
$\left((I-A)^{N+1}-\mathcal{P}_{\mathrm{Ker}(A_{\rho})}\right)h = h-\mathcal{P}_{\mathrm{Ker}(A_{\rho})}h - h_{N}$ where 
$h_{N}$ is the iterative approximation of $h$ in terms of Eq.(\ref{iterativesolution}), 
we see that
\begin{eqnarray}
\left|\left<h,f-\mathcal{P}_{\mathrm{Ker}(A_{\rho})}f - f_{N}\right>\right| \cr
 \qquad \leq \left\Vert h-\mathcal{P}_{\mathrm{Ker}(A_{\rho})}h \right\Vert_{L^{2}} \left\Vert f-\mathcal{P}_{\mathrm{Ker}(A_{\rho})}f - f_{N} \right\Vert_{L^{2}}
\end{eqnarray}
and
\begin{eqnarray}
\left|\left<h,f-\mathcal{P}_{\mathrm{Ker}(A_{\rho})}f - f_{N}\right>\right| \cr
 \qquad \leq \left\Vert f-\mathcal{P}_{\mathrm{Ker}(A_{\rho})}f \right\Vert_{L^{2}} \left\Vert h-\mathcal{P}_{\mathrm{Ker}(A_{\rho})}h - h_{N} \right\Vert_{L^{2}}.
\end{eqnarray}
By using $\left\Vert h-\mathcal{P}_{\mathrm{Ker}(A_{\rho})}h \right\Vert_{L^{2}} \leq \left\Vert h \right\Vert_{L^{2}}$ and 
setting $h:=\frac{1}{\mathrm{Volume}(U)}\chi_{{}_{U}}$ we have proved (i) and (iii).

Quite obviously, the same argument can be repeated with the projection operator
$\mathcal{P}_{\mathrm{Ker}(A_{\rho})}$ excluded from the equations, which proves (ii) and (iv).

Point (v) is proved by observing that for any $h\in L^{2}(X)$ one has 
$\left<h,f-f_{N}\right>=\left<h,(I-A)^{N+1}f\right>$, since $f-f_{N}=(I-A)^{N+1}f$. 
Due to the self-adjointness 
of the composite folding operator $A$, one has that 
$\left<h,f-f_{N}\right>=\left<(I-A)^{N+1}h,f\right>$. Since the identity 
$(I-A)^{N+1}h=h-h_{N}$ holds, one arrives at $\left<h,f-f_{N}\right>=\left<h-h_{N},f\right>$ 
and thus $\left\vert\left<h,f-f_{N}\right>\right\vert=\left\vert\left<h-h_{N},f\right>\right\vert$ is valid. 
Then, (v) is proved by simply substituting $h:=\xi_{{}_{U}}$.
\end{proof}

\vspace*{2mm}
\begin{remark}
The following remarks clarify the usability of Theorem~\ref{approximationerror}.
\begin{enumerate}[(i)]
\item By statement (i) and (ii) it is implied that the residual error averaged 
over a compact set $U\subset X$ scales as $\frac{1}{\sqrt{\mathrm{Volume}(U)}}$. 
In the language of histograms it means that it scales as one per square root of 
the histogram bin size.
\item The upper bounds (i), (iii) decrease monotonically to zero with increasing 
$N$. The upper bounds (ii) and (iv) decrease monotonically to the corresponding limits 
$\frac{1}{\sqrt{\mathrm{Volume}(U)}}\left\Vert \mathcal{P}_{\mathrm{Ker}(A_{\rho})}f\right\Vert_{L^{2}}$ and 
$\left\Vert f\right\Vert_{L^{2}} \left\Vert \mathcal{P}_{\mathrm{Ker}(A_{\rho})} \xi_{{}_{U}}\right\Vert_{L^{2}}$, 
respectively. Since $\left\Vert \xi_{{}_{U}} - \xi_{{}_{U,N}}\right\Vert_{L^{2}}$ 
is fully calculable, upper bound (iv) can be used to test whether the inverse 
of $A_{\rho}$ exists, i.e.\ whether $\mathcal{P}_{\mathrm{Ker}(A_{\rho})}=0$ holds, or if 
not, it may be used to quantify the contribution of the irrecoverable part $\mathcal{P}_{\mathrm{Ker}(A_{\rho})}f$.
\item Via spectral representation it is easy to see that $\left\Vert f_{N}\right\Vert_{L^{2}}$ 
converges to the limit $\left\Vert f-\mathcal{P}_{\mathrm{Ker}(A_{\rho})}f\right\Vert_{L^{2}}$ in a 
monotonically increasing way, i.e.\ may be used to approximate this unknown coefficient 
from below.
\item Again via using spectral representation, one can see that with 
fixed $N$ and $M>N$, the expressions 
$\left\Vert f_{M}-f_{N}\right\Vert_{L^{2}}$ and $\left\Vert \xi_{{}_{U,M}}-\xi_{{}_{U,N}}\right\Vert_{L^{2}}$ 
tend to the corresponding limits $\left\Vert f-\mathcal{P}_{\mathrm{Ker}(A_{\rho})}f - f_{N}\right\Vert_{L^{2}}$ and 
$\left\Vert \xi_{{}_{U}}-\mathcal{P}_{\mathrm{Ker}(A_{\rho})}\xi_{{}_{U}} - \xi_{{}_{U,N}}\right\Vert_{L^{2}}$ 
with increasing $M$, 
respectively, in a monotonically increasing way. Therefore, they can be used for 
approximation of these unknown coefficients from below.
\item As a consequence, the approximation error may be estimated for a fixed iteration 
order $N$ in the following way. 
For any $\varepsilon > 0$ there exists an iteration index threshold 
$M_{\varepsilon,N}>N$ such that for all $M>M_{\varepsilon,N}$ 
\begin{eqnarray}
\left\vert\frac{1}{\mathrm{Volume}(U)}\int_{x\in U} \left( f-\mathcal{P}_{\mathrm{Ker}(A_{\rho})}f - f_{N} \right)(x)\,\mathrm{d}x\right\vert \cr
\qquad \leq \frac{1}{\sqrt{\mathrm{Volume}(U)}}\,(1+\varepsilon)\left\Vert f_{M} - f_{N} \right\Vert_{L^{2}}
\end{eqnarray}
is valid. In addition, a closer, $U$-dependent estimate may be calculated: 
for any $\varepsilon > 0$ there exists an iteration index threshold 
$M_{\varepsilon,U,N}>N$ for which for 
all $M>M_{\varepsilon,U,N}$ the upper bound
\begin{eqnarray}
\left\vert\frac{1}{\mathrm{Volume}(U)}\int_{x\in U} \left( f-\mathcal{P}_{\mathrm{Ker}(A_{\rho})}f - f_{N} \right)(x)\,\mathrm{d}x\right\vert \cr
\qquad \leq (1+\varepsilon) \left\Vert f_{M}\right\Vert_{L^{2}} \, \left\Vert \xi_{{}_{U,M}} - \xi_{{}_{U,N}}\right\Vert_{L^{2}}
\end{eqnarray}
holds. Alternatively,
\begin{eqnarray}
\left\vert\frac{1}{\mathrm{Volume}(U)}\int_{x\in U} \left( f - f_{N} \right)(x)\,\mathrm{d}x\right\vert \cr
\qquad \leq (1+\varepsilon) \left\Vert f_{M}\right\Vert_{L^{2}} \, \left\Vert \xi_{{}_{U}} - \xi_{{}_{U,N}}\right\Vert_{L^{2}}
\end{eqnarray}
is also valid whenever $A_{\rho}$ is known to be one-to-one, which expression 
is slightly cheaper to calculate.
\item The identity (v) is particularly useful. In order to constructively 
evaluate it, one needs to use the fact that the sequence $(f_{N})_{N\in\mathbb{N}_{0}}$ 
converges to $f-\mathcal{P}_{\mathrm{Ker}(A_{\rho})}f$ in the $L^{2}$ sense. Thus, whenever 
$A_{\rho}$ is invertible, it converges to $f$ in the $L^{2}$ sense. In that case, 
the identity (v) can be rewritten as
\begin{eqnarray}
\left\vert\frac{1}{\mathrm{Volume}(U)}\int_{x\in U} \left( f - f_{N} \right)(x)\,\mathrm{d}x\right\vert \cr
\qquad = \lim_{M\rightarrow\infty} \left\vert \int \left(\xi_{{}_{U}}-\xi_{{}_{U,N}}\right)(x)\,f_{M}(x) \,\mathrm{d}x\right\vert.
\end{eqnarray}
Technically, the right side of this identity may be approximated by the integral 
$\left\vert \int \left(\xi_{{}_{U}}-\xi_{{}_{U,N}}\right)(x)\,f_{M}(x) \,\mathrm{d}x\right\vert$ 
with large enough $M$. For large $N$, even $M:=N$ may be used for evaluation 
of this expression.
\end{enumerate}
\label{approximationerrorrem}
\end{remark}

\subsection{\textbf{\textit{Estimation of statistical error}}}

Armed with the approximation error estimates of Theorem~\ref{approximationerror} one can construct 
penalty functions which define optimal stopping criterion of the iteration, 
and one can quantify the error of the approximation at finite iteration order 
which decreases with increasing iteration order.

In practice, however, the unfolding problem $g=A_{\rho}f + e$ may also contain 
a small statistical error term $e$ whose expectation value is zero, its exact 
value is unknown, but an estimate to the behavior of the random variable 
$e(x)$ for each $x\in X$ is available. Normally, the statistical 
covariance matrix $\mathrm{Cov}(e)$ is known along with the measured pdf 
$g$ and the known response function $\rho$. If, for instance, $g$ was a result 
of a measurement in the form of a histogram, 
then $\mathrm{Cov}(e)=\mathrm{Cov}(g)$ will be nothing but the diagonal 
matrix composed of the histogram bin entries. The question naturally arises: 
how can one quantify the propagated statistical error of the $N$-th iterative 
approximation of $f$, i.e.\ of $f_{N}$. In the followings we show an exact formula 
for the case when $g$ is measured as a histogram, i.e.\ when $g$ can be regarded as an 
$n$-component vector of real probability variables with known covariance.

\vspace*{2mm}
\begin{remark} The following simple facts in probability theory 
will aid the argumentation of the statistical error propagation.
\begin{enumerate}[(i)]
\item If $v$ is a $n$-component vector of real probability variables, then 
its covariance $\mathrm{Cov}(v)$ is an $n\times n$ real symmetric positive 
matrix. Therefore, for any $m\geq n$ there exists (not necessarily uniquely) a 
real $n\times m$ matrix $\mathrm{Err}(v)$ such that
\begin{eqnarray}
\mathrm{Cov}(v)=\mathrm{Err}(v) \mathrm{Err}(v)^{T}
\label{sqrt}
\end{eqnarray}
holds, the symbol $(\cdot)^{T}$ denoting matrix transpose. Indeed, because of 
realness, symmetricity and positivity of $\mathrm{Cov}(v)$ there exists uniquely 
a real symmetric positive $n\times n$ matrix satisfying Eq.(\ref{sqrt}), the 
square-root of $\mathrm{Cov}(v)$, and therefore 
$\mathrm{Err}(v)=\sqrt{\mathrm{Cov}(v)}$ may be chosen. 
Then, this may be extended to be $n\times m$ ($m\geq n$) by zeros without 
affecting Eq.(\ref{sqrt}). In some special cases, however, there also exists such 
$n\times m$ ($m\leq n$) real matrix $\mathrm{Err}(v)$ such that Eq.(\ref{sqrt}) still holds.
\item If $v$ is an $n$-component vector of real probability variables and 
$M$ is a real $m\times n$ matrix, then the standard error propagation formula
\begin{eqnarray}
\mathrm{Cov}(Mv) = M\mathrm{Cov}(v)M^{T}
\end{eqnarray}
holds.
\item As a consequence of the previous observations, one can express the 
standard error propagation formula also in the form
\begin{eqnarray}
\mathrm{Err}(Mv) = M\mathrm{Err}(v)
\end{eqnarray}
where $\mathrm{Err}(v)$ is any real $n\times n$ matrix satisfying Eq.(\ref{sqrt}),
and the resulting real $m\times n$ matrix $\mathrm{Err}(Mv)$ shall obey 
$\mathrm{Err}(Mv) \mathrm{Err}(Mv)^{T} = \mathrm{Cov}(Mv)$.
\item In our unfolding problem the $N$-th iterative approximation of $f$, i.e.\ $f_{N}$, 
may be expressed in the form
\begin{eqnarray}
f_{N} = \left(\sum_{n=0}^{N}\left(I-K_{\rho}^{-1}A_{\rho}^{T}A_{\rho}\right)^{n}\right)K_{\rho}^{-1}A_{\rho}^{T}g
\label{neumannseriesform}
\end{eqnarray}
which is manifestly linear in the measured pdf $g$. This fact may be used 
in order to construct statistical error propagation formula in terms of the 
previous observations.
\end{enumerate}
\label{statrem}
\end{remark}

\vspace*{2mm}
Armed with these equalities, we are ready to state the statistical error 
propagation formula for our unfolding method.

\vspace*{2mm}
\begin{theorem} (Statistical error) Take the iterative solution for the unfolding problem as in Eq.(\ref{iterativesolution}) and 
assume that the convergence conditions of Theorem~\ref{convergence} hold. 
Let $\mathrm{Cov}(g)$ be the $n\times n$ statistical covariance matrix of the 
measured pdf $g$, where $g$ is given in the form of an $n$-bin histogram. 
If $f$ and $f_{N}$ is modeled as an $m$-bin histogram, then the 
$m\times m$ covariance matrix of $f_{N}$, 
$\mathrm{Cov}(f_{N})$, may be obtained by the following iterative formula along 
with $f_{N}$:
\begin{eqnarray}
E_{0}   & := & K_{\rho}^{-1}A_{\rho}^{T}\mathrm{Err}(g),\cr
E_{N+1} & := & E_{N} +\left(E_{0} - K_{\rho}^{-1}A_{\rho}^{T}A_{\rho}E_{N} \right)\cr
        &    & (N\in \mathbb{N}_{0})
\label{eqstaterror}
\end{eqnarray}
where $E_{N}{E_{N}}^{T} = \mathrm{Cov}(f_{N})$ holds for each $N$.
\label{staterror}
\end{theorem}
\begin{proof}
This is a simple consequence of the linearity of the unfolding method Eq.(\ref{iterativesolution}), 
and of Remark~\ref{statrem}~(iv) combined with (iii) and then re-expressing it 
via iterative form.
\end{proof}

\vspace*{2mm}
\begin{remark}
The following remarks add some pieces of information about the practical usage 
of the statistical error propagation theorem.
\begin{enumerate}[(i)]
\item If the measured pdf $g$ is a histogram, then each component obeys 
Poisson distribution, and thus $\mathrm{Cov}(g)=\mathrm{diag}(g)$. Furthermore 
a real $n\times n$ matrix $\mathrm{Err}(g)$, satisfying $\mathrm{Err}(g)\mathrm{Err}(g)^{T}=\mathrm{Cov}(g)$, 
may be constructed by taking the componentwise square-root of $\mathrm{diag}(g)$. This 
can directly be used in calculation of $E_{0}$ in Theorem~\ref{staterror}.
\item If $f$ is modeled as a histogram with $m$ bins then for each iteration 
order $N$ the real matrix $E_{N}$ is of $m\times n$ type, i.e.\ 
$\mathrm{Cov}(f_{N})=E_{N}{E_{N}}^{T}$ shall be of $m\times m$ type.
\item The square-root of the diagonal elements of the covariance 
matrix $\mathrm{Cov}(f_{N})$ give the exact statistical errors of $f_{N}$ 
which then may be used to define an iteration stopping criterion, for instance 
the sum of the statistical errors may be required to be under a predefined 
threshold. 
One should not forget, however, that this unfolding method ---just as any 
other unfolding method--- introduces pretty strong correlations and thus 
the non-diagonal elements of $\mathrm{Cov}(f_{N})$ also play an important role when 
describing the characteristics of the statistical fluctuations of $f_{N}$.
\end{enumerate}
\end{remark}

\subsection{\textbf{\textit{Estimation of systematic error}}}

It was shown that in case of a statistical unfolding problem of the form 
$g = A_{\rho}f + e$ the quantification of the two competing error terms 
is possible: close upper bound to the convergent approximation error term was given, whereas 
exact error propagation formula to the divergent statistical error term was 
shown. A combination, such as the sum of these terms, may be considered as 
penalty function and the iteration may be stopped when the penalty function is minimal, 
furthermore these terms may be quantified at this optimal iteration order with the shown formulae.
In practice, however, one often faces the problem of systematic errors 
whenever the measured pdf contains some systematic distortion not 
accounted for in our model of response function, or equivalently, whenever our model 
of response function is slightly inaccurate. Formally we may write in such case 
that the actually measured pdf is $g+\delta{g} = A_{(\rho+\delta{\rho})}f + e$ 
where $\delta{\rho}$ is the deviation of the true response function 
$\rho+\delta{\rho}$ from our model response function $\rho$. Since by 
definition $g = A_{\rho}f + e$ would be the measured pdf in the absence of 
$\delta{\rho}$, one arrives at the relation $\delta{g}=A_{\delta{\rho}}f$ 
between $\delta{g}$ and $\delta{\rho}$. When applying the iterative solution 
Eq.(\ref{iterativesolution}) using $\rho$ to the actually measured pdf $g+\delta{g}$, the 
$N$-th iterative estimate of the true unknown pdf $f$ shall contain a propagated 
contribution $\delta{f}_{N}$ which needs to be quantified. In experimental 
practice, the systematic error of the actually measured pdf is given in terms 
of some close upper estimate $s{g}$ for which $|\delta{g}|\leq s{g}$ holds, 
or similarly as a close upper estimate $s{\rho}$ for which $|\delta{\rho}|\leq s{\rho}$ is valid. 
Our aim is to provide some upper estimate to $|\delta{f}_{N}|$ based on $s{g}$ or 
$s{\rho}$, for any given iteration order $N\in \mathbb{N}_{0}$. For this, let
us introduce the following normalization factors
\begin{eqnarray}
C_{\rho,s{g}} := \sqrt{\int \left( K_{\rho}^{-1}A_{\rho}^{T}s{g} \right)^{2}(x)\,\mathrm{d}x}
\end{eqnarray}
if the systematic errors are known in terms of $s{g}$, and
\begin{eqnarray}
D_{\rho,s{\rho}} := \sqrt{\sup_{x\in X} \int \int \left(K_{\rho}^{-1}A_{\rho}^{T}s{\rho}\right)(y\vert z)\, \left(K_{\rho}^{-1}A_{\rho}^{T}s{\rho}\right)(y\vert x)\, \mathrm{d}y\, \mathrm{d}z}
\end{eqnarray}
if the systematic errors are known in terms of $s{\rho}$.

\vspace*{2mm}
\begin{theorem} (Systematic error) Take the iterative solution for the unfolding problem as in Eq.(\ref{iterativesolution}) and 
assume that the conditions of convergence hold. Then, the following upper bounds 
are valid on the systematic deviation $\delta{f}_{N}$ of the $N$-th iterative 
approximation of $f$, $f_{N}$.
\begin{enumerate}[(i)]
\item For the average of $\delta{f}_{N}$ over any compact set $U\subset X$ one has
\begin{eqnarray}
\left\vert\frac{1}{\mathrm{Volume}(U)}\int_{x\in U} \delta{f}_{N}(x)\,\mathrm{d}x\right\vert 
 \leq \left\Vert \Xi_{{}_{U,N}}\right\Vert_{L^{2}}\,C_{\rho,s{g}}
\end{eqnarray}
where $\xi_{{}_{U}}:=\frac{1}{\mathrm{Volume}(U)}\chi_{{}_{U}}$ and 
$\Xi_{{}_{U,N}}$ is defined by the iteration
\begin{eqnarray}
\Xi_{{}_{U,0}}   & := & \xi_{{}_{U}}, \cr
\Xi_{{}_{U,N+1}} & = & \Xi_{{}_{U,N}} + \left( \Xi_{{}_{U,0}} - K_{\rho}^{-1}A_{\rho}^{T}A_{\rho} \Xi_{{}_{U,N}} \right) \cr
                   &   & (N\in \mathbb{N}_{0}).
\end{eqnarray}
\item Alternatively,
\begin{eqnarray}
\left\vert\frac{1}{\mathrm{Volume}(U)}\int_{x\in U} \delta{f}_{N}(x)\,\mathrm{d}x\right\vert 
 \leq \left\Vert \Xi_{{}_{U,N}}\right\Vert_{L^{2}}\,D_{\rho,s{\rho}}\left\Vert f\right\Vert_{L^{2}}.
\end{eqnarray}
\item The upper bound
\begin{eqnarray}
\left\vert\frac{1}{\mathrm{Volume}(U)}\int_{x\in U} \delta{f}_{N}(x)\,\mathrm{d}x\right\vert 
 \leq \int \left\vert K_{\rho}^{-1}A_{\rho}\Xi_{{}_{U,N}}\right\vert(y)\, s{g}(y) \,\mathrm{d}y
\end{eqnarray}
also holds.
\item Alternatively,
\begin{eqnarray}
\left\vert\frac{1}{\mathrm{Volume}(U)}\int_{x\in U} \delta{f}_{N}(x)\,\mathrm{d}x\right\vert 
 \leq \int \left(K_{\rho}^{-1}A_{s{\rho}}^{T}\left\vert A_{\rho}\Xi_{{}_{U,N}}\right\vert\right)(x)\, \left\vert f\right\vert(x) \,\mathrm{d}x.
\label{eqsystsrho}
\end{eqnarray}
\item More specifically,
\begin{eqnarray}
\left\vert\frac{1}{\mathrm{Volume}(U)}\int_{x\in U} \delta{f}_{N}(x)\,\mathrm{d}x\right\vert 
 \leq \left\Vert f\right\Vert_{L^{1}} \; \sup_{x\in X}\left(K_{\rho}^{-1}A_{s{\rho}}^{T}\left\vert A_{\rho}\Xi_{{}_{U,N}}\right\vert\right)(x).
\end{eqnarray}
Here, whenever $f$ was a pdf, then $\left\Vert f\right\Vert_{L^{1}}=1$ automatically holds.
\end{enumerate}
\label{systematicerror}
\end{theorem}
\begin{proof} We begin the proof by recalling that because of Eq.(\ref{neumannseriesform}) 
and its modified form
\begin{eqnarray}
f_{N}+\delta{f}_{N} = \left(\sum_{n=0}^{N}\left(I-K_{\rho}^{-1}A_{\rho}^{T}A_{\rho}\right)^{n}\right)K_{\rho}^{-1}A_{\rho}^{T}(g+\delta{g})
\end{eqnarray}
in presence of systematic distortions, we have that
\begin{eqnarray}
\delta{f}_{N} = \left(\sum_{n=0}^{N}\left(I-K_{\rho}^{-1}A_{\rho}^{T}A_{\rho}\right)^{n}\right)K_{\rho}^{-1}A_{\rho}^{T}\delta{g}
\end{eqnarray}
holds, where $\delta{g}$ is the unaccounted systematic distortion of the measured pdf, 
which is related to the unaccounted systematic distortion of the response function 
$\delta{\rho}$ by $\delta{g}=A_{\delta{\rho}}f$.

Again, we use the notation $A:=K_{\rho}^{-1}A_{\rho}^{T}A_{\rho}$ and use its 
spectral representation as in the proof of Theorem~\ref{convergence}. With this, 
one has
\begin{eqnarray}
\left<h,\delta{f}_{N}\right> = \int_{\lambda\in[0,1]} 1\,\sum_{n=0}^{N}(1-\lambda)^{n}\,\mathrm{d}\left<h,P(\lambda)K_{\rho}^{-1}A_{\rho}^{T}\delta{g}\right>
\end{eqnarray}
for any $h\in L^{2}(X)$. From that, using Remark~\ref{generalizedcauchyschwarz} 
we arrive at
\begin{eqnarray}
\left|\left<h,\delta{f}_{N}\right>\right| \cr
 \qquad \leq \sqrt{\int_{\lambda\in[0,1]} \left|\sum_{n=0}^{N}(1-\lambda)^{n}\right|^{2} \,\mathrm{d}\left<h,P(\lambda)h\right>} \cr
 \qquad \qquad \sqrt{\int_{\lambda\in[0,1]} |1|^{2}\,\mathrm{d}\left<K_{\rho}^{-1}A_{\rho}^{T}\delta{g},P(\lambda)K_{\rho}^{-1}A_{\rho}^{T}\delta{g}\right>} \cr
 \qquad = \left\Vert\sum_{n=0}^{N}(I-A)^{n}h\right\Vert_{L^{2}} \, \left\Vert K_{\rho}^{-1}A_{\rho}^{T}\delta{g}\right\Vert_{L^{2}} \cr
 \qquad = \left\Vert H_{N}\right\Vert_{L^{2}} \, \left\Vert K_{\rho}^{-1}A_{\rho}^{T}\delta{g}\right\Vert_{L^{2}}
\label{systproof1}
\end{eqnarray}
where the notation $H_{N}:=\sum_{n=0}^{N}(I-A)^{n}h$ was introduced. 
It is quite evident that $H_{N}$ may be calculated using the iterative form
\begin{eqnarray}
H_{0}   & := & h, \cr
H_{N+1} & := & H_{N} + \left(H_{0} - A H_{N}\right) \cr
        &    & (N\in\mathbb{N}_{0})
\end{eqnarray}
in order to evaluate $\left\Vert H_{N}\right\Vert_{L^{2}}$.

An upper bound for $\left\Vert K_{\rho}^{-1}A_{\rho}^{T}\delta{g}\right\Vert_{L^{2}}$ may be readily constructed using the inequality
\begin{eqnarray}
\left\Vert K_{\rho}^{-1}A_{\rho}^{T}\delta{g}\right\Vert_{L^{2}}^{2} \leq \left\Vert K_{\rho}^{-1}A_{\rho}^{T}s{g}\right\Vert_{L^{2}}^{2} = C_{\rho,s{g}}^{2}
\end{eqnarray}
which is seen to hold using Fubini's theorem and monotonicity of integration, 
where non-negativity of $\rho$ and $s{g}$ is tacitly assumed as previously.

Now, by setting $h:=\xi_{{}_{U}}$, part (i) of the theorem is proved.

Part (ii) may be proved by using the relation $\delta{g}=A_{\delta{\rho}}f$ 
which implies that
\begin{eqnarray}
\left\Vert K_{\rho}^{-1}A_{\rho}^{T}\delta{g}\right\Vert_{L^{2}}^{2} = \left\Vert K_{\rho}^{-1}A_{\rho}^{T}A_{\delta{\rho}}f\right\Vert_{L^{2}}^{2} \leq \left\Vert K_{\rho}^{-1}A_{\rho}^{T}A_{s{\rho}}f\right\Vert_{L^{2}}^{2}
\end{eqnarray}
again because of Fubini's theorem and monotonicity of integration, where one 
should note that $\rho$, $s{\rho}$ and $f$ is assumed to be non-negative as 
previously. Then, we see that
\begin{eqnarray}
\left\Vert K_{\rho}^{-1}A_{\rho}^{T}A_{s{\rho}}f\right\Vert_{L^{2}}^{2} & = & \left<f,K_{\rho}^{-1}A_{s{\rho}}^{T}A_{\rho}K_{\rho}^{-1}A_{\rho}^{T}A_{s{\rho}}f\right> \cr
 & \leq & \left\Vert f\right\Vert_{L^{2}}^{2} \left\Vert K_{\rho}^{-1}A_{s{\rho}}^{T}A_{\rho}K_{\rho}^{-1}A_{\rho}^{T}A_{s{\rho}} \right\Vert_{L^{2}\rightarrow L^{2}}
\end{eqnarray}
holds. Realizing that the $L^{2}$ operator norm of the positive self adjoint operator 
$K_{\rho}^{-1}A_{s{\rho}}^{T}A_{\rho}K_{\rho}^{-1}A_{\rho}^{T}A_{s{\rho}}$ 
can be bound via the Riesz-Thorin theorem similarly as for $K_{\rho}^{-1}A_{\rho}^{T}A_{\rho}$
in proof of Theorem~\ref{convergence} we conclude that the pertinent operator 
norm is bound by $D_{\rho,s{\rho}}^{2}$.

Part (iii) is proved by using the self-adjointness of $A$ and that the adjoint 
of $A_{\rho}^{T}$ is $A_{\rho}$. Due to that, for any $h\in L^{2}(X)$, one has
\begin{eqnarray}
\left<h,\delta{f}_{N}\right>=\left<K_{\rho}^{-1}A_{\rho}H_{N},\delta{g}\right>
\label{systl2}
\end{eqnarray}
with the previous notations. Due to the monotonicity of integration, then the 
identity 
$\left\vert\left<h,\delta{f}_{N}\right>\right\vert\leq\left<\left\vert K_{\rho}^{-1}A_{\rho}H_{N}\right\vert, s{g}\right>$ 
is obtained, since $\vert\delta{g}\vert\leq s{g}$ holds. 
When setting $h:=\xi_{{}_{U}}$ and correspondingly 
$H_{N}:=\Xi_{{}_{U,N}}$, this is nothing but (iii).

Part (iv) is proved by using Eq.(\ref{systl2}) and $\delta{g}=A_{\delta{\rho}}f$, 
furthermore that the adjoint of $A_{\delta{\rho}}$ is $A_{\delta{\rho}}^{T}$. 
With that, one has
$\left<h,\delta{f}_{N}\right>=\left<K_{\rho}^{-1}A_{\delta{\rho}}^{T}A_{\rho}H_{N},f\right>$. 
Using $\left\vert\delta{\rho}\right\vert\leq s{\rho}$ and the monotonicity of integration, one arrives at 
$\left\vert\left<h,\delta{f}_{N}\right>\right\vert\leq\left<K_{\rho}^{-1}A_{s{\rho}}^{T}\left\vert A_{\rho}H_{N}\right\vert,\left\vert f\right\vert\right>$. 
The upper bound (iv) is obtained, whenever $h:=\xi_{{}_{U}}$ and $H_{N}:=\Xi_{{}_{U,N}}$ is set.

Part (v) is a consequence of (iv), applying H\"older's inequality, in addition.
\end{proof}

\vspace*{2mm}
\begin{remark}
The following remarks provide some more explanation about the usability of the 
above results on upper estimation of the systematic errors of $f_{N}$ originating 
from the systematic errors of the measured pdf $g$ or of the response function $\rho$.
\begin{enumerate}[(i)]
\item For any given iteration order $N\in\mathbb{N}_{0}$ the upper estimate 
(i) of Theorem~\ref{systematicerror} bounds the systematic deviation of the 
unfolded pdf $f_{N}$ averaged over any compact set, in a manifestly calculable 
way if the systematic errors of the measured pdf are given. In the language of histograms 
this means that bin-by-bin upper bound to the systematic error of the unfolded 
pdf is available in terms of the systematic error of the measured pdf.
\item The upper estimate (ii) of Theorem~\ref{systematicerror} provides an 
alternative bound for the same quantity for the case when the systematic 
errors are known in terms of the systematic error of the response function. This, 
similarly to Theorem~\ref{approximationerror}~(iv), needs the unknown value of $\Vert f\Vert_{L^{2}}$ 
which may be circumvented in the analogy of Remark~\ref{approximationerrorrem}~(v). Namely, 
for any $\varepsilon>0$ there exists an iteration 
index threshold $M_{\varepsilon}\in\mathrm{N}_{0}$ such that for all $M>M_{\varepsilon}$ one has
\begin{eqnarray}
\left\vert\frac{1}{\mathrm{Volume}(U)}\int_{x\in U} \delta{f}_{N}(x)\,\mathrm{d}x\right\vert \cr
 \qquad \leq \left\Vert \Xi_{{}_{U,N}}\right\Vert_{L^{2}}\,D_{\rho,s{\rho}}\,(1+\varepsilon)\left\Vert f_{M}\right\Vert_{L^{2}}
\end{eqnarray}
whenever $A_{\rho}$ is one-to-one, because then in the light of Remark~\ref{approximationerrorrem}~(iii), 
$\left\Vert f_{M}\right\Vert_{L^{2}}$ as a function of $M$ converges to $\left\Vert f\right\Vert_{L^{2}}$ in a monotonically increasing way.
\item The right side of Eq.(\ref{eqsystsrho}) may be approximated by 
\begin{eqnarray}
\int \left(K_{\rho}^{-1}A_{s{\rho}}^{T}\left\vert A_{\rho}\Xi_{{}_{U,N}}\right\vert\right)(x)\, \left\vert f_{M}\right\vert(x) \,\mathrm{d}x
\end{eqnarray}
due to $\vert f\vert=f$ and because $f_{M}$ converges to $f$ as $M\rightarrow\infty$ 
in the $L^{2}$ sense, whenever $A_{\rho}$ is invertible. 
For large $N$, the approximative formula with $M:=N$ may be used.
\end{enumerate}
\end{remark}

\section{Generalization to the context of probability measures}
\label{measures}

In rare cases one faces the problem that the distributions in question 
cannot be described in terms of pdfs, only in terms of probability measures 
instead.\footnote{A measure is a set function of the subsets of the probability base space. A common example of measures is the Dirac delta.} 
Such practical cases may arise for instance when the folding operator 
represents kinematics of particle decays \cite{laszlo2006}. Therefore, 
it is interesting to ask the question whether the iterative unfolding method 
Eq.(\ref{iterativesolution}) applies in the framework of probability measures.

\vspace*{2mm}
\begin{remark} Let us recall some notions in measure theory \cite{dinculeanu1967}.
\begin{enumerate}[(i)]
\item A complex measure $F$ over $X$ is a complex valued $\sigma$-additive 
set function on the Borel $\sigma$-algebra of the subsets of $X$. The variation 
of the complex measure $F$ is the non-negative valued measure $|F|$ defined 
by the requirement: for a Borel set $E$ the value of $|F|(E)$ is the 
supremum of $\sum_{k=0}^{K}|F(E_{k})|$ for any splitting $E_{1},\dots,E_{K}$ of $E$, 
i.e.\ for all such finite system of disjoint Borel sets $E_{1},\dots,E_{K}$ whose 
union totals up to $E$. The measures with finite variation, i.e.\ 
which have $|F|(X)<\infty$, form a Banach space with the norm being 
$\Vert F\Vert := |F|(X)$. We shall denote this space by $M(X)$.
\item A probability measure $F$ on $X$ is a non-negative measure on the 
Borel $\sigma$-algebra of $X$ with the requirement $F(X)=1$. Thus, quite 
naturally, a probability measure on $X$ resides in $M(X)$.
\end{enumerate}
\end{remark}

\vspace*{2mm}
We continue with the formal definition of folding operators whose response 
function is described by a measure rather than a function.

\vspace*{2mm}
\begin{definition}
A mapping $Q:\,X\rightarrow M(Y), x\mapsto Q(\cdot|x)$ is called folding measure 
if for every $x\in X$ the measure $Q(\cdot|x)$ is a non-negative measure on $Y$ 
with $Q(Y|x)=1$ (i.e.\ $Q(\cdot|x)$ is a probability measure for all $x\in X$),
and for every Borel set $E$ in $Y$ the function $x\mapsto Q(E|x)$ is measurable.
\end{definition}

\vspace*{2mm}
\begin{remark}
A possible usual generalization is when inefficiencies are also allowed, i.e.\ the 
less restrictive condition $Q(Y|x)\leq 1$ is required for all $x\in X$. The 
results throughout this paper also holds for that case.
\end{remark}

\vspace*{2mm}
It follows from the definition that a folding measure $Q$ may be viewed as a 
conditional probability measure over the product space $Y\times X$. Quite 
evidently, if $\rho$ is a response function then 
$Q_{\rho}(E|x):=\int_{y\in E}\rho(y|x)\,\mathrm{d}y$ 
defines a folding measure.

\vspace*{2mm}
\begin{definition}
Let $Q$ be a folding measure. Then, the linear map
\begin{eqnarray}
A_{Q}:\,M(X)\rightarrow M(Y),\, F\mapsto \left(\int Q(\cdot|x)\,\mathrm{d}F(x)\right)
\end{eqnarray}
is called the folding operator by $Q$.
\end{definition}

\vspace*{2mm}
\begin{remark} The remarks below follow from the definition \cite{laszlo2006}.
\begin{enumerate}[(i)]
\item A folding operator $A_{Q}$ is well-defined as for all points $x\in X$ and 
Borel sets $E$ of $Y$ the inequality $Q(E|x)\leq 1$ holds, thus the function 
$x\mapsto Q(E|x)$ is integrable by any measure with finite variation.
\item The monotonicity of integration implies that a folding operator is 
continuous and $\Vert A_{Q}\Vert_{M(X)\rightarrow M(Y)} = 1$, just as in the case of $L^{1}$ theory. 
If inefficiencies are allowed, $\Vert A_{Q}\Vert_{M(X)\rightarrow M(Y)} \leq 1$ holds.
\item The folding operators defined by folding measures is a generalization 
of the folding operators by response functions.
\end{enumerate}
\end{remark}

\vspace*{2mm}
As in the $L^{1}$ theory, the convolutions represent an important class 
of folding operators.

\vspace*{2mm}
\begin{definition}
A folding operator $A_{Q}$ is called a convolution if its folding measure 
is translationally invariant in the sense that $Y=X$ and for all $x,z\in X$ 
and Borel sets $E$ one has $Q(E|x+z)=Q(E-z|x)$.
\end{definition}

\vspace*{2mm}
\begin{remark} The followings are important properties of convolution operators with measures \cite{laszlo2006}.
\begin{enumerate}[(i)]
\item Whenever the folding operator $A_{Q}$ by a folding measure $Q$ is a 
convolution, $Q$ may be expressed by a single probability measure 
$R:=Q(\cdot|0)$ in the form of $Q(E|x)=R(E-x)$ 
for all $x\in X$ and Borel set $E$. The alternative notation $R\star F:=A_{Q}F$ 
is often used in such case ($F\in M(X)$). Note that the convolution is commutative, 
i.e.\ one has $R\star F=F\star R$ for all $R,F\in M(X)$.
\item Fourier transformation of measures in $M(X)$ can also be defined and has similar 
properties as in the $L^{1}$ case, except that the Fourier transform functions 
do not decay at infinity, i.e.\ the Riemann-Lebesgue lemma does not hold. 
Only the boundedness of Fourier transforms are guaranteed.
\item Properties of convolution operators are similarly related to the Fourier 
transform of the underlying probability measure, as in the $L^{1}$ theory. For instance, a convolution operator 
is one-to-one if and only if its Fourier transform is nonzero almost everywhere.
\item It is easily seen that if $\varphi\in L^{1}(X)$ and $F\in M(X)$, then 
$\varphi\star F$ is a function in $L^{1}(X)$. Combining this with Remark~\ref{rieszthorincor}~(ii) 
we conclude that if $\varphi\in L^{p}(X)\cap L^{1}(X)$ 
then for all $F\in M(X)$ the function $\varphi\star F\in L^{p}(X)\cap L^{1}(X)$ ($1\leq p\leq\infty$). 
That is, probability measures may be mapped into pdfs in $L^{p}(X)$ via convolution by 
a pdf integrable on the $p$-th power.
\end{enumerate}
\end{remark}

\vspace*{2mm}
Armed with the introduced notions we may try to ask the question whether one 
can generalize the results in Section~\ref{lineariterativeunfolding} to probability measures.

\vspace*{2mm}
\begin{remark} The following results are generalization of the results in Section~\ref{lineariterativeunfolding} for probability measures.
\begin{enumerate}[(i)]
\item The naive application of Neumann series fails to work similarly as 
in the $L^{1}$ framework. This is because as proved in \cite{laszlo2006} one has 
$\left\Vert I-A_{Q}\right\Vert_{M(X)\rightarrow M(X)} = 2$ whenever $Q(\{y\}|y)=0$ for any point $y$ 
--- which is the generic case.
\item The convergence and error propagation results of Theorem~\ref{convergence}, 
\ref{approximationerror}, \ref{staterror}, \ref{systematicerror} may be generalized in a 
similar manner to Remark~\ref{improvement}~(i)-(ii). Namely, instead of the 
original problem $G=A_{Q}F$ one may consider the modified version 
$\eta\star G=A_{\eta\star Q}F$ to be solved for $F$, where $\eta$ is a 
square-integrable pdf whose Fourier transform is nowhere vanishing. 
In this case, the folding operator $A_{Q}$ is mapped to be a folding 
operator by a response function $A_{\eta\star Q}$ instead, 
as we have $\eta\star A_{Q}F = A_{\eta\star Q}F$ for any $F\in M(X)$. 
Furthermore, for each $x\in X$ the pdf $\eta\star Q(\cdot|x)$ is 
square-integrable. Then, the iteration
\begin{eqnarray}
K_{\eta\star Q} & := & \sup_{x\in X} \int \int (\eta\star Q)(y\vert z)\, (\eta\star Q)(y\vert x)\, \mathrm{d}y\, \mathrm{d}\mu(z), \cr
F_{0}    & := & K_{\eta\star Q}^{-1} A_{\eta\star Q}^{T} \;\eta\star G, \cr
F_{N+1}  & := & F_{N} + \left(F_{0} - K_{\eta\star Q}^{-1}A_{\eta\star Q}^{T}A_{\eta\star Q}F_{N}\right) \cr
         &    & (N\in \mathbb{N}_{0}).
\label{iterativesolution3}
\end{eqnarray}
obeys the very same convergence and error propagation properties as 
stated in Theorem~\ref{convergence}, \ref{approximationerror}, \ref{staterror}, \ref{systematicerror}, 
whenever $K_{\eta\star Q}<\infty$ and when the unknown probability measure $F$ 
corresponds to a square-integrable pdf with respect to some a priori given 
non-negative valued measure $\mu$ over $X$. This latter requirement means that 
$F=f\mu$ needs to be satisfied with some non-negative measure $\mu$ over $X$ 
and with some $\mu$-measurable function $f:X\rightarrow \mathbb{R}_{0}^{+}$ 
for which $\int \left\vert f\right\vert^{2}(x) \,\mathrm{d}\mu(x) < \infty$ 
needs to hold.
\end{enumerate}
\end{remark}

\vspace*{2mm}
The previous observations conclude that whenever the unknown distribution is 
described by a pdf which is square-integrable with respect to some volume 
measure, then the folding measure may be conditioned in a way that the 
iterative unfolding Eq.(\ref{iterativesolution}) applies to it.

\section{The discrete case}
\label{discretecase}

For better illustration, we specialize our results in Section~\ref{lineariterativeunfolding} and \ref{measures} 
to the case when the unknown probability distribution along with the response 
function and the measured probability distribution is discrete. In that case 
the measured pdf $g$ and the unknown pdf $f$ is a finite dimensional vector 
of non-negative entries, and the folding operator $A_{\rho}$ is simply a finite 
dimensional matrix with non-negative entries as well. Our equation to solve 
is then the matrix equation $g=A_{\rho}f$ for $f$, or in case of presence of 
measurement errors $e$, the matrix equation $g=A_{\rho}f+e$. We also assume 
that the entries of $f$, $A_{\rho}$ and $g$ are probabilities, i.e.\ they are normalized 
such that $\sum_{i}g_{i}=1$, $\sum_{i}f_{i}=1$ and $\sum_{j}\left(A_{\rho}\right)_{ji}=1$, 
or $\sum_{j}\left(A_{\rho}\right)_{ji}\leq1$ in case of presence of inefficiencies.

Then, the iterative solution of our discrete unfolding problem reads as
\begin{eqnarray}
K_{\rho} & := & \max_{i} \sum_{j} \sum_{k} \left(A_{\rho}\right)_{ji}\,\left(A_{\rho}\right)_{jk}, \cr
f_{0}    & := & K_{\rho}^{-1} A_{\rho}^{T} g, \cr
f_{N+1}  & := & f_{N} + \left(f_{0} - K_{\rho}^{-1}A_{\rho}^{T}A_{\rho}f_{N}\right) \cr
         &    & (N\in \mathbb{N}_{0}).
\label{discreteiterativesolution}
\end{eqnarray}
where $A_{\rho}^{T}$ is the matrix transpose of $A_{\rho}$. A simple observation 
shows that Eq.(\ref{discreteiterativesolution}) is nothing but an iterative 
form of
\begin{eqnarray}
K_{\rho} & := & \max_{i} \sum_{j} \sum_{k} \left(A_{\rho}\right)_{ji}\,\left(A_{\rho}\right)_{jk}, \cr
f_{N}    & := & \sum_{n=0}^{N} (I-K_{\rho}^{-1}A_{\rho}^{T}A_{\rho})^{n}K_{\rho}^{-1}A_{\rho}^{T}g \cr
         &    & (N\in \mathbb{N}_{0}).
\end{eqnarray}
$I$ denoting the identity matrix. Due to the results of 
Section~\ref{lineariterativeunfolding} and \ref{measures}, the convergence 
of this approximation is monotonic in the $l^{2}$ vector norm, and also holds 
entrywise, however with possibly quite different convergence rates for 
different vector entries. Along with this, all the convergence and error propagation properties 
listed in Section~\ref{lineariterativeunfolding} and \ref{measures} hold, 
independently of the fineness of the discretization. This decoupling from the discretization 
is quite important, as it shows that in the presented method the discretization 
does not become an important ingredient of the regularization procedure itself 
in case when $f$, $g$ and $\rho$ are in reality continuum distributions, 
modelled and measured as histograms.

\section{Numerical example}
\label{numericalexample}

In this section the performance of the proposed method is illustrated on a 
numerical example. The example calculation is implemented via the C library 
\texttt{libunfold} \cite{laszlo2011}, also including the automatic approximation, statistical 
and systematic error propagation formulae presented in the paper. The shown 
example is also shipped with the pertinent library. The illustrative case 
was deliberately chosen in a way when the response function is not translationally 
invariant, i.e.\ when ordinary deconvolution methods are not sufficient.

\begin{figure}[!ht]
\begin{minipage}{\textwidth}
\begin{minipage}{\textwidth}
\includegraphics[width=6.3cm]{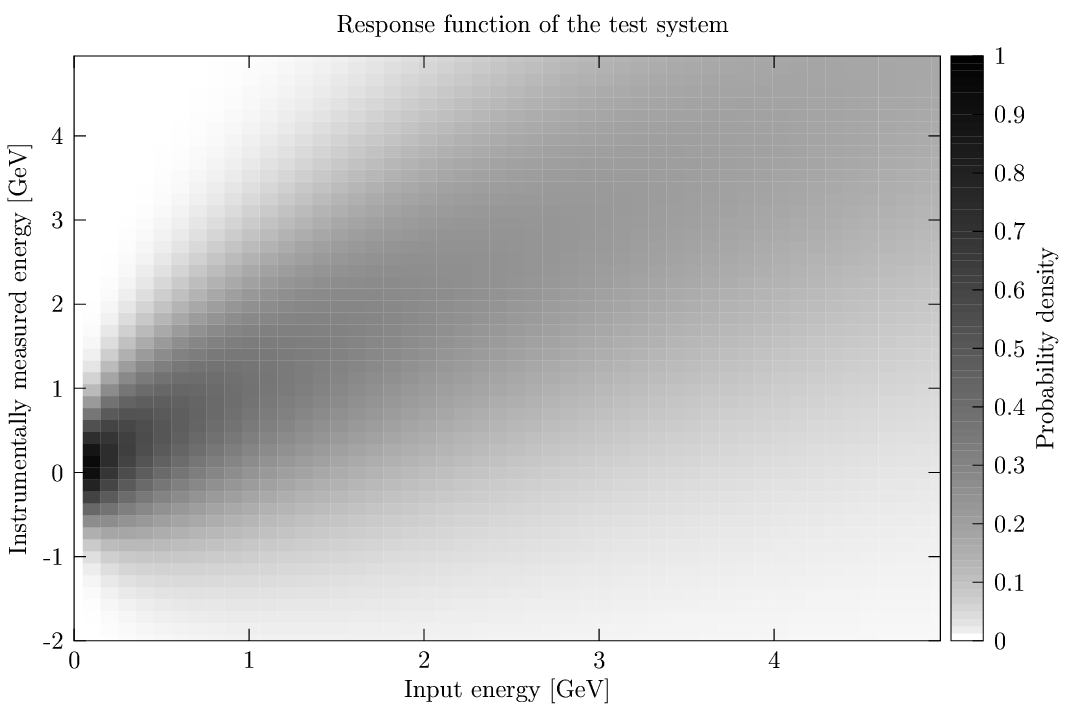}\includegraphics[width=6.3cm]{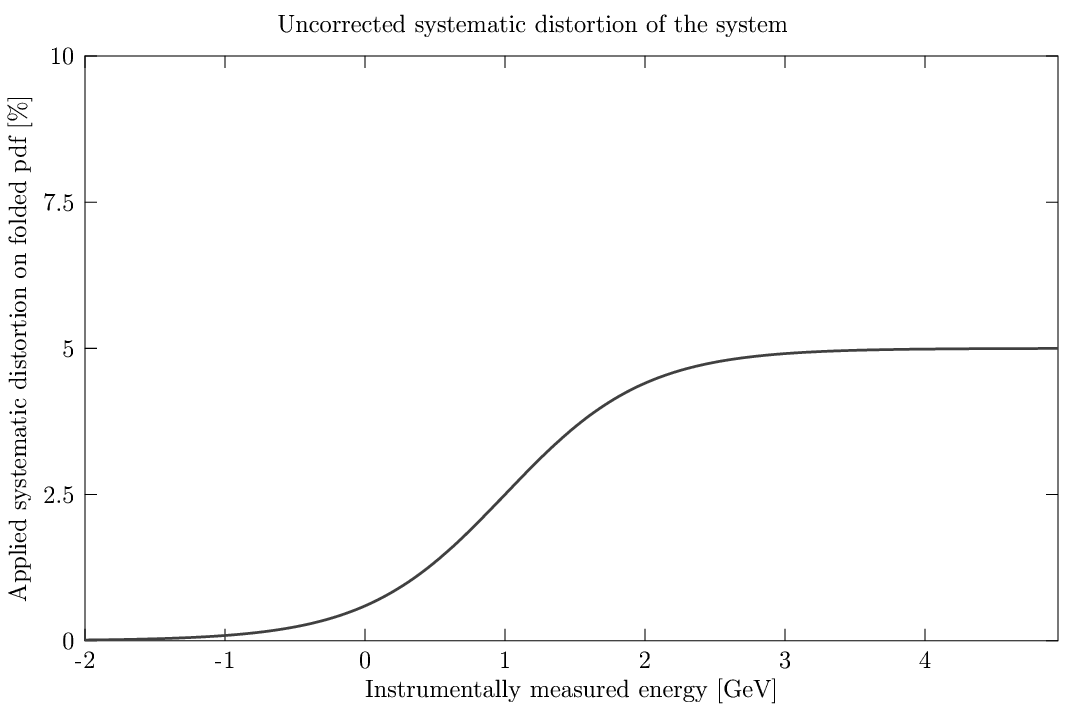}
\end{minipage}

\begin{minipage}{\textwidth}
\vspace*{-35mm}
\includegraphics[width=6.3cm]{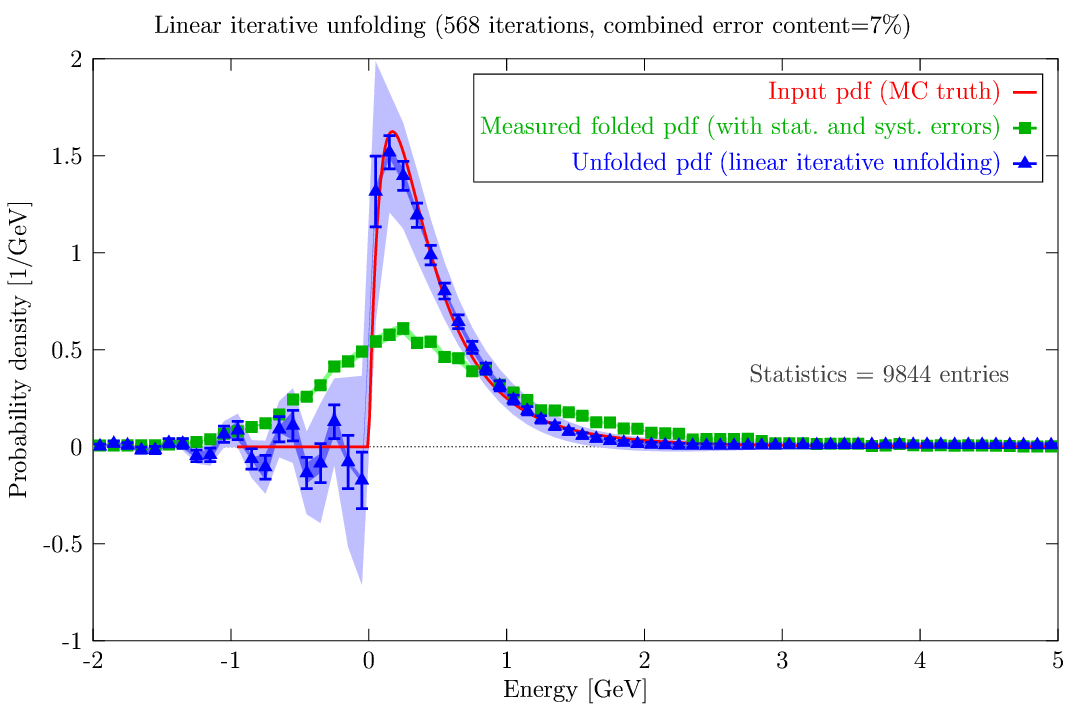}\includegraphics[width=6.3cm]{fig/cmshcal_errorcontent.eps}
\end{minipage}
\end{minipage}
\caption{(Color online) Top left: illustration of the response function of our test 
example. The color intensity indicates the probability density of the response function. 
Top right: illustration of the unaccounted systematic distortion applied 
to the folded pdf in our test example. The solid curve indicates the systematic 
distortion (an inefficiency, in our example) on the unfolded pdf, which is 
assumed to be not exactly quantifiable, and therefore is not corrected for in the simulated measured 
pdf. Only an upper bound for the systematic distortion, called to be the 
systematic error, is assumed to be known for the simulated measured pdf. That 
is taken to be a constant $5\%$ upper bound in the example. 
Bottom left: the true input pdf (solid line), the simulated measured pdf (squares) 
and the unfolded pdf (triangles) by the proposed method. The pdfs are shown 
together with their bin-by-bin statistical errors (error bars), systematic 
errors (error bands), and approximation errors (narrow error bands). 
Bottom right: evolution of the bin-by-bin maximum of the approximation error 
(circles), statistical error (diamonds), and systematic error (flipped triangles) 
as a function of the number of iterations. Note that the binwise approximation 
errors converge to zero, but not in a monotonic manner, which 
explains the slight increase of that term after about $300$ iterations. If 
the iteration was continued, that term indeed converged to zero, but with 
several local minima, i.e.\ ``waves'' or ``jumps'' are seen in the convergence 
curve. On the other hand, the binwise statistical and systematic error term 
are seen simply to diverge, as expected. The competition of these three 
error terms gives a possibility to define a stopping criterion.}
\label{figunfolding}
\end{figure}

Our simulated measurement scenario is the following. We would like to measure 
the true pdf of a quantity, namely of the energy of produced charged 
particles in a high energy particle collision experiment. This true pdf 
used in our toy Monte Carlo shall be a parametrization of a real measurement at 
the LHC accelerator \cite{khachatryan2010} at CERN. It is of the form
\begin{eqnarray}
 E & \mapsto & f(E):=\chi_{{}_{[0,\infty[}}(E)\;\;\vert E\vert\,\frac{(n-1)(n-2)}{(n\,T)^{2}}\,\left(1+\frac{\vert E\vert}{n\,T}\right)^{-n}
\label{mcinput}
\end{eqnarray}
with parameters $n=6.6$ and $T=0.145\,\mathrm{GeV}$.
The response function 
\begin{eqnarray}
(E_{\mathrm{measured}},\, E_{\mathrm{true}}) & \mapsto & \rho(E_{\mathrm{measured}}\,\vert\, E_{\mathrm{true}})
\label{mcresponse}
\end{eqnarray}
shall be such a cpdf that for each fixed value $E_{\mathrm{true}}>0$ the pdf 
\begin{eqnarray}
E_{\mathrm{measured}} & \mapsto & \rho(E_{\mathrm{measured}}\,\vert\, E_{\mathrm{true}})
\end{eqnarray}
shall be a Gaussian pdf with a mean of $E_{\mathrm{true}}$ and standard deviation 
of $a+\sqrt{b\,E_{\mathrm{true}}}+c\,E_{\mathrm{true}}$, with parameter values 
$a=0.150\,\mathrm{GeV}$, $b=0.7174\,\mathrm{GeV}$, $c=0.074$. This response function 
models the behavior of a calorimeter device used for the energy measurement of 
particles, namely of the HCAL calorimeter \cite{yazgan2009} of the CMS experiment 
at the LHC accelerator at CERN. In the simulated measurement scenario 
$10^{4}$ Monte Carlo samples according to the pdf Eq.(\ref{mcinput}) was 
generated, and its corresponding smeared response according to 
Eq.(\ref{mcresponse}) was generated. These responses were assumed to be collected 
with an inefficiency of
\begin{eqnarray}
E_{\mathrm{measured}} & \mapsto & \frac{1}{2}\left(1+\tanh\left(\frac{E_{\mathrm{measured}}-\mathcal{E}}{\Delta}\right)\right)\,d
\label{mcineff}
\end{eqnarray}
with parameters $\mathcal{E}=1\,\mathrm{GeV}$, $\Delta=1\,\mathrm{GeV}$ and $d=0.05$, 
i.e.\ with an inefficiency not greater than $5\%$ on the overall measurement 
domain. The collected responses were histogramed, providing the measured pdf $g$ 
with our non-ideal detector. By construction, the statistical covariance 
matrix of the histogram $g$ shall be $\mathrm{diag}(g)$. 
The inefficiency profile Eq.(\ref{mcineff}) causing a systematic deviation of the 
measured pdf from the folded pdf by Eq.(\ref{mcresponse}), is assumed 
to be not known quantitatively and therefore is not corrected for. It is assumed, however, 
that an overall $5\%$ upper bound to this systematic deviation is known, 
being the systematic error of the measured pdf, i.e.\ one has 
$sg=0.05\,g$. 
With these inputs, the linear iterative unfolding according to 
Eq.(\ref{iterativesolution}) was performed. 
The approximation errors were quantified using Remark~\ref{approximationerrorrem}~(vi).
The propagated statistical errors 
were calculated according to Eq.(\ref{eqstaterror}). 
The propagated systematic errors were quantified using Theorem~\ref{systematicerror}~(iii). 
The iteration was stopped 
when the combined statistical, approximation and systematic error exceeded a 
predefined threshold of $7\%$. The result of the numerical test is shown in Fig.~\ref{figunfolding}. 
Note, that more optimalized stopping criteria can also be invented, using the 
estimates for the approximation error, statistical error and systematic error. 
A natural candidate can be a double-threshold criterion: the approximation 
error needs to be below a threshold (sufficient shape restoration), whereas 
the combined statistical and systematic error must stay below an upper bound 
(divergence regularization). Also, the iteration might be stopped at the 
error optimum: at the minimum of the combined approximation, statistical and 
systematic error. Note, however, that one often might require a better shape 
reconstruction at the expense of increased statistical and systematic errors, 
as also seen in the shown example.

\section{Concluding remarks}
\label{conclusions}

In this paper we presented mathematical proofs of convergence and error propagation 
formulae for a linear iterative unfolding method \cite{laszlo2012} in the probability theory context.
It was shown that the pertinent method is convergent in the `binwise' sense 
under quite generic conditions, which does hold in case of many practical applications. 
Furthermore, explicit formulae for the three important error terms, the 
approximation error, the statistical error and the systematic errors were derived. 
These can be used to define optimal iteration stopping criterion and 
quantification of errors therein. The key element of the proofs is the Riesz-Thorin theorem 
mapping the original $L^1$ problem to $L^2$ with a subsequent usage of 
spectral theory of $L^2$ operators. The typical use-cases of the method are those 
unfolding problems which cannot be handled by statistical deconvolution 
\cite{dattner2011, dattner2016}, due to the absence of translational invariance 
of the response function. The possibility for propagation of the systematic 
errors is a special advantage, which deserves to be emphasized for experimental 
applications.

The pertinent method is also available as a C numerical library \cite{laszlo2011}. 
Using that, the method was demonstrated on a numerical example. 
The algorithm could be included in the ROOUnfold package \cite{adye2011} in the future, or 
in the GNU Scientific Library \cite{gsl}.

The present paper can serve also as a good motivation to perform similar 
convergence and error propagation studies on an other iterative 
unfolding method \cite{dagostini1995, zech2013, richardson1972, lucy1974, shepp1982, kondor1983, multhei1987mma, multhei1987nim}, 
also called the method of convergent weights or iterative Bayesian unfolding. 
That method is non-linear and therefore is somewhat more complicated 
to study, however can be more suitable for unfolding problems in probability 
theory as it conserves the integral and non-negativity of probability 
density functions. Although widely used and numerically very promising, 
so far little is known on the convergence properties 
of that algorithm, and nothing is known about its error propagation. Our 
proposed method can be considered as the ``linearized'' version of that 
method, and thus the presented results are expected to provide clues also 
for the convergence and error propagation properties of the method of 
convergent weights or iterative Bayesian unfolding.

\section*{Acknowledgments}

The author would like to thank to Tam\'as Matolcsi for valuable comments and 
for reading various versions of the manuscript, furthermore to Dezs\H{o} Varga for discussions on 
the physical applications and on the relevance of error propagation formulae, 
in particular for the systematic errors. 
This work was supported in part by the Momentum (`Lend\"ulet') program of the 
Hungarian Academy of Sciences under the grant number LP2013-60. 
The author would also like to acknowledge the support of the 
J\'anos Bolyai Research Scholarship of the Hungarian Academy of Sciences.


\begin{thebibliography}{99}
\bibitem{laszlo2012} {\it A.~L\'aszl\'o}: A linear iterative unfolding method. \textit{J.~Phys.~Conf.~Ser.} {\bf 368} (2012), 012043.
\bibitem{laszlo2006} {\it A.~L\'aszl\'o}: A robust iterative unfolding method for signal processing. \textit{J.~Phys.} {\bf A39} (2006), 13621. Zbl 1107.94004
\bibitem{cowan2002} {\it G.~Cowan}: \textit{Proceedings of Conference on Advanced Statistical Techniques in Particle Physics (18-22 March 2002, Durham, United Kingdom)} (2002, IPPP/02/39, Durham) p 248
\bibitem{blobel2008} {\it V.~Blobel}: Unfolding for HEP Experiments (\textit{Talk at DESY Computing Seminar}, 2008) \texttt{http://www.desy.de/\~{}blobel/DESYcompsem08.pdf}
\bibitem{zech2010} {\it G.~Bohm, G.~Zech}: Introduction to Statistics and Data Analysis for Physicists (2010, Hamburg: Verlag Deutsches Elektronen-Synchrotron).
\bibitem{kuusela2015} {\it M.~Kuusela, V.~M.~Panaretos}: Statistical unfolding of elementary particle spectra: empirical Bayes estimation and bias-corrected uncertainty quantification. {\it Ann.~Appl.~Stat.} {\bf 9} (2015), 1671.
\bibitem{kuusela2016} {\it M.~Kuusela, P.~B.~Stark}: Shape-constrained uncertainty quantification in unfolding steeply falling elementary particle spectra. {\it Preprint} (2016) [\texttt{arXiv:1512.00905}].
\bibitem{dembinski2013} {\it H.~P.~Dembinski, M.~Roth}: An algorithm for automatic unfolding of one-dimensional distributions. \textit{Nucl.~Instr.~Meth.} {\bf A729} (2013), 410.
\bibitem{dattner2011} {\it I.~Dattner, A.~Goldenshluger, A.~Juditsky}: On deconvolution of distribution functions. \textit{Ann.~Stat.} {\bf 39} (2011), 2477. Zbl 1232.62056
\bibitem{dattner2016} {\it I.~Dattner, M.~Rei{\ss}, M.~Trabs}: Adaptive quantile estimation in deconvolution with unknown error distribution. \textit{Bernoulli} {\bf 22} (2016), 143. Zbl 06543266
\bibitem{fan1991} {\it J.~Fan}: On the optimal rates of convergence for nonparametric deconvolution problems. \textit{Annals~of~Stat.} {\bf 19} (1991), 1257. Zbl 0729.62033
\bibitem{hesse2006} {\it C.~H.~Hesse}: Iterative density estimation from contaminated observations. \textit{Metrika} {\bf 64} (2006), 151. Zbl 1100.62042
\bibitem{lacour2013} {\it F.~Comte, C.~Lacour}: Anisotropic adaptive kernel deconvolution. \textit{Ann.~Inst.~H.~Poincar\'e Prob.~Stat.} {\bf 49} (2013), 569. Zbl 06171260
\bibitem{liu1989} {\it M.~C.~Liu, R.~L.~Taylor}: A consistent nonparametric density estimator for the deconvolution problem. \textit{Can.~J.~Stat.} {\bf 17} (1989), 427. Zbl 0694.62017
\bibitem{stefanski1990} {\it L.~A.~Stefanski, R.~J.~Carol}: Deconvoluting kernel density estimators. \textit{Statistics} {\bf 21} (1990), 169. Zbl 0697.62035
\bibitem{kalifa2003} {\it J.~Kalifa, B.~Rouge}: Deconvolution by Thresholding in Mirror Wavelet Bases. \textit{IEEE~Trans.~on~Image~Proc.} {\bf 12} (2003), 446.
\bibitem{hoecker1996} {\it A.~Hoecker, V.~Kartvelishvili}: SVD Approach to data unfolding. \textit{Nucl.~Instr.~Meth.} \textbf{A372} (1996), 469.
\bibitem{dagostini1995} {\it G.~D'Agostini}: A multidimensional unfolding method based on Bayes' theorem. \textit{Nucl.~Instr.~Meth.} \textbf{A362} (1995), 487.
\bibitem{zech2013} {\it G.~Zech}: Iterative unfolding with the Richardson-Lucy algorithm. \textit{Nucl.~Instr.~Meth.} \textbf{A716} (2013), 1.
\bibitem{laszlo2008} {\it C.~Alt et al}: High Transverse Momentum Hadron Spectra at $\sqrt{s_{{}_{NN}}}=17.3\,\mathrm{GeV}$, in Pb+Pb and p+p Collisions. \textit{Phys.~Rev.} \textbf{C77} (2008), 034906.
\bibitem{richardson1972} {\it W.~H.~Richardson}: Bayesian-based iterative method of image restoration. {\it J.~Opt.~Soc.~of~Amer.} {\bf A62} (1972), 55.
\bibitem{lucy1974} {\it L.~B.~Lucy}: An iterative technique for the rectification of observed distributions. {\it Astronomical~Journal} {\bf 79} (1974), 745.
\bibitem{shepp1982} {\it L.~A.~Shepp, Y.~Vardi}: Maximum likelihood reconstruction for emission tomography. {\it IEEE~Trans.~Med.~Imag.} {\bf 1} (1982), 113.
\bibitem{kondor1983} {\it A.~Kondor}: Method of convergent weights -- An iterative procedure for solving Fredholm's integral equations of the first kind. {\it Nucl. Instr. Meth.} {\bf 216} (1983), 177.
\bibitem{multhei1987mma} {\it H.~N.~M\"ulthei, B.~Schorr}: On an iterative method for a class of integral equations of the first kind. \textit{Math.~Meth.~Appl.~Sci.} \textbf{9} (1987), 137. Zbl 0628.65130
\bibitem{multhei1987nim} {\it H.~N.~M\"ulthei, B.~Schorr}: On an iterative method for the unfolding of spectra. \textit{Nucl.~Instr.~Meth.} \textbf{A257} (1987), 371.
\bibitem{lax2002} {\it P.~D.~Lax}: Functional analysis (2002, Chichester: Wiley-Interscience). Zbl 1009.47001
\bibitem{rudin1973} {\it W.~Rudin}: Functional Analysis (1973, New York: McGraw-Hill). Zbl 0253.46001
\bibitem{arfken2013} {\it G.~Arfken}: Fourier Convolution theorem 20.4 \textit{Mathematical Methods for Physicists} 7rd edn (2013, Amsterdam: Elsevier/Academic Press) pp 985. Zbl 1239.00005
\bibitem{bracewell1999} {\it P.~Bracewell}: Convolution theorem \textit{The Fourier Transform and Its Applications} 3rd edn (1999, New York: McGraw-Hill) pp 108.
\bibitem{landweber1951} {\it L.~Landweber}: An iteration formula for Fredholm integral equations of the first kind. \textit{Am.~J.~Math.} \textbf{73} (1951), 615. Zbl 0043.10602
\bibitem{folland1999} {\it G.~B.~Folland}: Real Analysis: Modern Techniques and Their Applications 2nd edn (1999, Wiley-Interscience). Zbl 0924.28001
\bibitem{dinculeanu1967} {\it N.~Dinculeanu}: Vector Measures (1967, Elsevier). Zbl 0142.10502
\bibitem{laszlo2011} {\it A.~L\'aszl\'o}: The libunfold package (2011, \textit{Source code}) \newline\texttt{http://www.rmki.kfki.hu/\~{}laszloa/downloads/libunfold.tar.gz}
\bibitem{khachatryan2010} {\it V.~Khachatryan et al (the CMS collaboration)}: Transverse-Momentum and Pseudorapidity Distributions of Charged Hadrons in pp Collisions at $\sqrt{s}=7$ TeV. \textit{Phys.~Rev.~Lett.} {\bf 105} (2010), 022002.
\bibitem{yazgan2009} {\it E.~Yazgan (for the CMS collaboration)}: The CMS barrel calorimeter response to particle beams from 2 to 350 GeV/c. \textit{J.~Phys.~Conf.~Ser.} {\bf 160} (2009), 012056.
\bibitem{adye2011} {\it T.~Adye et al}: The ROOUnfold package \newline\texttt{http://hepunx.rl.ac.uk/\~{}adye/software/unfold/RooUnfold.html}
\bibitem{gsl} The GNU Scientific Library: \texttt{http://www.gnu.org/software/gsl}
\end{thebibliography}
\end{document}